\documentclass[usenatbib]{mn2e}
\newcommand{\bib}{\bibitem[\protect\citeauthoryear}
\newcommand{\pa}{\,\rlap{\raise 0.5ex\hbox{$\propto$}}{\lower 1.0ex\hbox{$\sim$}}\,}
\usepackage[dvips]{graphicx}
\begin{document}
\title[FR\,I jet spectra]
{The spectra of jet bases in FR\,I radio galaxies: implications for particle acceleration}
\author[R.A. Laing et al.]
   {R.A. Laing \thanks{E-mail: rlaing@eso.org}$^{1}$, 
    A.H. Bridle$^2$\\
    $^1$ European Southern Observatory, Karl-Schwarzschild-Stra\ss e 2, D-85748 
    Garching-bei-M\"unchen, Germany \\
    $^2$ National Radio Astronomy Observatory, 520 Edgemont Road, Charlottesville,
    VA 22903-2475, U.S.A.}

\date{Received }
\maketitle

\begin{abstract}
We present accurate, spatially resolved imaging of radio spectra at
the bases of jets in eleven low-luminosity (Fanaroff-Riley I) radio
galaxies, derived from Very Large Array (VLA) observations.  We pay
careful attention to calibration and to the effects of random and
systematic errors, and we base the flux-density scale on recent
measurements of VLA primary amplitude calibrators by \citet{PB}.  We
show images and profiles of spectral index over the frequency range
1.4 -- 8.5\,GHz, together with values integrated over fiducial regions
defined by our relativistic models of the jets.  We find that the
spectral index $\alpha$, defined in the sense $I(\nu) \propto
\nu^{-\alpha}$, decreases with distance from the nucleus in all of the
jets. The mean spectral indices are $0.66 \pm 0.01$ where the jets
first brighten abruptly and $0.59 \pm 0.01$ after they recollimate.
The mean {\sl change} in spectral index between these locations, which
is independent of calibration and flux-density scale errors and is
therefore more accurately and robustly measured, is $-0.067 \pm
0.006$. Our jet models associate the decrease in spectral index with a
bulk deceleration of the flow from $\approx 0.8c$ to $\la 0.5c$.  We
suggest that the decrease is the result of a change in the
characteristics of ongoing particle acceleration.  One possible
acceleration mechanism is the first-order Fermi process in mildly
relativistic shocks: in the Bohm limit, the index of the electron
energy spectrum, $p = 2\alpha+1$, is slightly larger than 2 and
decreases with velocity upstream of the shock.  This possibility is
consistent with our measurements, but requires shocks throughout the
jet volume rather than at a few discrete locations.  A second
possibility is that two acceleration mechanisms operate in these jets:
one (with $p = 2.32$) dominant close to the galactic nucleus and
associated with high flow speeds, another (with $p = 2.18$) taking
over at larger distances and slower flow speeds.
\end{abstract}


\section{Introduction}
\label{Introduction}

Jets from active galactic nuclei (AGN), microquasars and gamma-ray bursts
can accelerate electrons over enormous ranges of energy, generating
electromagnetic radiation over the entire observable spectrum
(e.g.\ \citealt{Mkn501}).  They may even produce the protons and
heavier nuclei which make up the highest-energy cosmic rays
\citep{Auger}.  The mechanisms by which jets accelerate particles to
relativistic energies are poorly understood, however. The form
of the electron energy spectrum inferred from observations of their
synchrotron emission carries information about acceleration
and loss processes. The optically-thin synchrotron spectrum observed
at radio frequencies typically has an approximately power-law form
$I(\nu) \propto \nu^{-\alpha}$, with spectral index $\alpha \approx
0.5 - 0.75$.  This corresponds to a particle energy spectrum $n(E)dE = n_0
E^{-p}dE$ with $p = 2\alpha+1 \approx 2$ -- 2.5.  At low energies
where loss processes are unimportant, the slope of this spectrum 
should be characteristic of the acceleration mechanism.  For
example, first-order Fermi acceleration in strong, non-relativistic
shocks \citep{Bell78,BO78} has $p = 2$ ($\alpha = 0.5$), whereas the
asymptotic value of $p$ for ultrarelativistic shocks is 2.23
\citep{Kirk00,LP03}.  Mildly relativistic shocks can produce a variety
of spectral slopes \citep{SB12}.  Other mechanisms which have been 
suggested for electron acceleration in jets include shear \citep{SO,RD04,RD06},
second-order Fermi acceleration by turbulence
(e.g.\ \citealt{Petrosian12}) and magnetic reconnection
(e.g.\ \citealt{BL00}).  The dominant acceleration mechanism may change 
both locally and globally with position in the jets, for example if the flow 
decelerates from supersonic to subsonic, entrains clumps of denser material,
develops significant shear or becomes turbulent.  Accurate,
spatially-resolved measurements of the synchrotron spectra from AGN
jets can, in principle, differentiate between acceleration mechanisms and 
show us which processes dominate in different regions of the jets.

The kiloparsec-scale bases of jets in Fanaroff-Riley Class\,I radio
galaxies (FR\,I; \citealt{FR74}) are known to emit at all frequencies from 
the radio to the soft X-ray regimes.  The integrated spectra of the jet
bases can usually be fit by broken power laws with $\alpha \approx 0.5 - 0.7$ 
at low frequencies,
steepening to $\alpha \approx 0.7 - 1.4$ in the X-ray band
\citep*{Evans05,H01,H05,Lanz,PW05,Perlman10,Worrall10,Harwood}.
Imaging of linear polarization (e.g.\ \citealt{Perlman99,Perlman10})
confirms that the emission mechanism at optical frequencies is indeed
synchrotron radiation, as is well established for the radio band.  The
X-ray spectrum is also in accord with this interpretation.  The
corresponding energy distribution (also a broken power law) is often
referred to as a ``single electron population'' in the sense that it
is continuous between the presumed, but usually unobservable,
lower and upper limits.  Observations of the closest radio galaxies,
Cen\,A and M\,87 \citep{H07,Worrall08,PW05}, show that there are both
small-scale inhomogeneities and large-scale gradients in the
broad-band spectra.

An important implication of the detection of extended X-ray emission
from AGN jets is that ongoing, distributed particle acceleration is
necessary to counterbalance synchrotron cooling. The inferred
synchrotron lifetimes of the radiating electrons are typically
hundreds of years, corresponding to light-travel distances similar to,
or smaller than, the jet radii.  The acceleration mechanism(s) must
therefore act over much of the length of a jet, although not
necessarily over its entire volume \citep{PW05}.  Acceleration cannot
be restricted to one or two discrete locations such as jet-crossing
shocks.

Precise synchrotron spectral indices at radio frequencies were
determined for the jets in two FR\,I sources by
\citet{Young}. Combining these results with others from the
literature, they found $\langle\alpha\rangle = 0.55$, apparently
inconsistent with the value of $\alpha = 0.5$ ($p = 2.0$) predicted
for first-order Fermi acceleration at strong, non-relativistic shocks.

We are studying the kinematics and dynamics of 
jets in FR\,I radio galaxies
by fitting
deep, high-resolution VLA observations with model brightness
distributions for symmetrical, decelerating relativistic flows
\citep{LB02a,CL,CLBC,LCBH06,LB12}.  In the course of this work, we
have acquired well calibrated and uniformly imaged data at multiple frequencies for 
the jets in ten FR\,I radio galaxies.  These observations, supplemented 
with additional data from the VLA Archive, are ideal for the study
of jet base spectra over the frequency range 1.4 -- 8.5\,GHz at high
spatial resolution and we have published spectral-index images for
many of the sources elsewhere \citep{LCBH06,LCCB06,3c31ls,LGBPB}.
Data of similar quality, reduced in a similar way, are also available for 
the FR\,I radio galaxy 3C\,449 whose jets are too close to the plane 
of the sky to model using our methods, but which was studied for 
different reasons by \citet{Guidetti10}.
 
Our initial work on the jet spectra in 3C\,31, NGC\,315 and 3C\,296
(summarized in \citealt{Girdwood}) confirmed \citet{Young}'s result
that the jet base spectral indices can be significantly higher than
0.5 and consistently found a value of $\alpha \approx 0.62$ near
the flaring points where the jets brighten abruptly. In a detailed
study of the spectrum of the brighter jet in the giant radio galaxy
NGC\,315 \citep{LCCB06}, we found that the jet spectral index
decreases with distance from the AGN,
contrary to the expectations of simple synchrotron-loss models, and
that there is significant transverse spectral structure, with a higher
spectral index (steeper spectrum) on the jet axis than at its edges.

\begin{table}
\caption{Source names, redshifts and linear scales. (1) name as used
  in this paper; (2) alternative names; (3) redshift; (4) linear scale
  in kpc\,arcsec$^{-1}$; (5) reference for
  redshift. \label{tab:sources}}
\begin{tabular}{lllll}
\hline
&&&&\\
Source & Alternate & $z$ & scale & Ref \\
NGC\,193 & UGC\,408, PKS\,0036+03  &0.0147 &0.300 &4 \\
NGC\,315 &                         &0.0165 &0.335 &6 \\
3C\,31   & NGC\,383                &0.0170 &0.346 &5 \\
0206+35 & UGC\,1651, 4C\,35.03 &0.0377 &0.748 &2 \\
0326+39 &                      &0.0243 &0.490 &3 \\
0755+37 & NGC\,2484            &0.0428 &0.845 &4 \\
3C\,270     & NGC\,4261            &0.0075 &0.154 &6 \\
M\,84       & 3C\,272.1, NGC\,4374 &0.0035 &0.073 &6 \\
3C\,296     & NGC\,5532            &0.0247 &0.498 &3 \\
1553+24 &                      &0.0426 &0.841 &1 \\
3C\,449     &                      &0.0171 &0.347 &1 \\
&&&&\\
\hline
\end{tabular}
References: 1 \citet{RC3}; 2 \citet{Falco99}; 3 \citet{Miller02}; 4 \citet{Ogando}; 5 \citet{Smith2000}; 6 \citet{Trager}. 
\end{table}

In the present paper, our aims are:
\begin{enumerate}
\item to present a coherent analysis of the jet base spectra for the
  whole sample, with careful attention to the flux-density scale
  \citep{PB}, calibration errors, noise measurement and correction for
  contaminating emission from surrounding structures;
\item to establish whether the flattening of the spectrum with
  increasing distance from the AGN that we found in earlier work is
  typical of FR\,I jets;
\item to measure the spectral indices at physically meaningful
  locations defined by the geometry, velocity and proper emissivity of
  the jets, to determine the dispersion between sources and to search
  for relations between spectrum and flow parameters;
\item to compare these measured values of spectral index with the
  predictions of particle-acceleration models.
\end{enumerate}
Section~\ref{Obs-red} presents the sample, gives references to
observations and basic data reduction (including a discussion of the
flux-density scale) and describes the additional image processing we
have performed to derive the final spectral-index
images. Section~\ref{Models} outlines our jet-modelling method and
defines the standard regions for which we tabulate integrated spectral
indices.  In Section~\ref{alphas} we present the total-intensity and
spectral-index images, show profiles along the jet axes and tabulate
integrated values at fiducial locations.  We discuss the results in
Section~\ref{discuss} and summarize our conclusions in
Section~\ref{summary}. Appendix~\ref{0326_1553} outlines the data
reduction for archival observations of two sources which have not been
presented elsewhere and Appendix~\ref{notes} gives notes on individual
sources, primarily to document instrumental issues and to reference
earlier work on spectral imaging.

\section{Observations and data reduction}
\label{Obs-red}

\subsection{The sources}
\label{Sources}

Our modelling technique requires us to be able to image the jets on
both sides of the nucleus with good transverse resolution at high
signal-to-noise ratio, and for their intensity ratio close to the
nucleus to be $\ga$5:1 (Section~\ref{Models}).  This led to the
selection of ten sources. We have added 3C\,449, which satisfies the
first criterion, but not the second. The sources are listed in
Table~\ref{tab:sources}, together with their redshifts and the
corresponding linear scales for a concordance cosmology with
$\Omega_{\rm M} = 0.3$, $\Omega_\Lambda = 0.7$ and H$_0$ =
70\,km\,s$^{-1}$\,Mpc$^{-1}$.

\begin{table*}
\caption{Parameters of the VLA images used to derive spectral
  indices. (1) source name; (2) resolution (FWHM), in arcsec; (3)
  rotation of image with respect to the sky, in deg; (4) centre
  frequency, in MHz; (5) VLA configurations used to make the image (H denotes
  a hybrid configuration); (6) correction factor applied to published
  observations to bring them onto the flux-density scale of
  \citet{PB}; (7) off-source noise level, in
  $\mu$Jy\,beam$^{-1}$; (8) noise level close to the core, in
  $\mu$Jy\,beam$^{-1}$; (9) reference for image; (10) method used to
  derive spectral-index images and averaged values (2 for
  two-frequency; PL for a power-law fit). Cols\,11 -- 13 give the
  parameters for lobe subtraction, in arcsec. (11) minimum width of
  reference region; (12) maximum width; (13) smoothing
  kernel. Cols\,11 - 13 are blank if no lobe subtraction was
  done. \label{tab:images}}
\begin{tabular}{llrlllllllrrr}
\hline
Source   & FWHM   &PA& Freq & Conf &Corr & $\sigma_{\rm off}$    & $\sigma_{\rm core}$    & Reference & Fit & Min & Max & Smooth \\
         & arcsec &deg& (MHz)     &      & & \multicolumn{2}{c}{$\mu$Jy\,b$^{-1}$} &     &     &\multicolumn{3}{c}{arcsec}\\
&&&&&&&&&&&\\
NGC\,193 & 1.6    &175.2& 4860.1    & ABCD & 0.991& 7.5 & 15  & \citet{LGBPB} & 2   &12&18&1.5\\
         &        && 1365.0    & ABC  & 1.029& 36  & 72  &               &     &&&\\
&&&&&&&&&&&\\
NGC\,315 & 5.5    &$-41.5$& 4985.1    & ABCD & 0.955& 15  & 41  & \citet{LCCB06}& PL  &&&\\
         &        && 1665.0    & BC   & 1.019& 37  &120  &               &     &&&\\
         &        && 1485.0    & BC   & 1.023& 37  & 37  &               &     &&&\\
         &        && 1413.0    & ABC  & 1.026& 38  & 46  &               &     &&&\\
         &        && 1365.0    & BC   & 1.026& 41  & 41  &               &     &&&\\
&&&&&&&&&&&\\
3C\,31   & 1.5    &$-70.3$& 8440.0    & ABCD & 0.956& 11  & 18  & \citet{3c31ls} & PL  &&&\\
         &        && 4985.0    & BCD  & 0.964& 12  & 22  &               &     &&&\\
         &        && 1636.0    & ABCD & 1.013& 72  & 150 &               &     &&&\\
         &        && 1485.0    & ABCD & 1.018& 73  & 150 &               &     &&&\\
         &        && 1435.0    & ABCD & 1.020& 78  & 110 &               &     &&&\\
         &        && 1365.0    & ABCD & 1.023& 76  & 130 &               &     &&&\\
&&&&&&&&&&&\\
0206+35  & 1.2    &$-41.0$& 4860.1    & BC   & 0.988& 12  & 120 & \citet{LGBPB}  & 2   &9&10&1.0\\
         &        && 1425.0    & AB   & 1.026& 19  & 210 &               &     &&&\\
&&&&&&&&&&&\\
0326+39  & 1.5    &3.0& 8460.1    & ABCD & 0.967& 6.9 & 14  & \citet{CL}    & PL  &&&\\
         &        && 4885.1    & HCD  & 0.980& 22  & 51  & Appendix~\ref{0326_1553} &    &&&\\
         &        && 1425.0    & ABCD & 1.025& 44  & 99  &               &     &&&\\
&&&&&&&&&&&\\
0755+37  & 1.3    &158.5& 4860.1    & BCD  & 0.980& 7.8 & 13  &  \citet{LGBPB} & 2   &30&45&3.0\\
         &        && 1425.0    & ABC  & 1.022& 20  & 36  &               &     &&&\\
&&&&&&&&&&&\\
3C\,270  & 1.65   &3.5& 4860.1    & ABCD & 0.982& 21  & 37  & Laing, Guidetti \& & 2   &14.4&18&2.0\\
         &        && 1388.5    & ABC  & 1.022& 21  & 50  & Bridle, in preparation   &     &&&\\
&&&&&&&&&&&\\
M\,84    & 1.65   &$-88.0$& 4860.1    & ABC  & 0.980& 15  & 39  &  \citet{LGBPB} & 2   &6&9&1.5\\
         &        && 1413.0    & AB   & 1.022& 140 & 150 &               &     &&&\\ 
&&&&&&&&&&&\\
3C\,296  & 1.5    &$-125.9$& 8460.1    & ABCD & 0.971& 9.5 & 9.5 & \citet{LCBH06}& 2   &17.7&19.2&1.5\\
         &        && 1406.75   & ABCD & 1.021& 27  & 48  &               &     &&&\\
&&&&&&&&&&&\\
1553+24  & 1.5    &$-46.0$& 8460.1    & ABC  & 0.974& 8.5 & 13  & \citet{CL}    & PL  &&&\\
         &        && 4860.1    & BCD  & 0.980& 22  & 30  &  Appendix~\ref{0326_1553} &    &&&\\
         &        && 1385.1    & ABC  & 1.023& 35  & 42  &                    &    &&&\\
&&&&&&&&&&&\\
3C\,449  & 1.25   &80.4& 8385.1    & ABCD & 0.987& 14  & 24  & \citet{Guidetti10}& PL &&&\\
         &        && 4985.1    & ABCD & 0.982& 18  & 28  & \citet{Feretti99} &    &&&\\
         &        && 4685.1    & ABCD & 0.983& 17  & 27  &                   &    &&&\\
         &        && 1485.0    & ABCD & 1.021& 35  & 45  &                   &    &&&\\
         &        && 1465.0    & ABCD & 1.021& 48  & 58  &                   &    &&&\\
         &        && 1445.0    & ABCD & 1.021& 21  & 31  &                   &    &&&\\
         &        && 1365.0    & ABCD & 1.023& 37  & 47  &                   &    &&&\\
&&&&&&&&&&&\\
\hline
\end{tabular}
\end{table*}

\subsection{VLA images}
\label{Images}

\subsubsection{Observations and references}
\label{obsref}

The observations were made using various combinations of frequencies
and array configurations in the 1.3 -- 1.7\,GHz, 4.5 -- 5.0\,GHz and
8.3 -- 8.5\,GHz bands of the VLA. We have published comprehensive
descriptions of the observations and data reduction for almost all of
the images used in this paper. The only exceptions are the
lower-frequency images for two sources, 0326+39 and 1553+24. These
have not been published elsewhere and the reduction of VLA archive
observations is described in Appendix~\ref{0326_1553}.  All of the
images used in the present paper are listed in Table~\ref{tab:images},
with references to descriptions of the observations and reductions.
We summarize some general points below.

\subsubsection{Calibration and imaging}
\label{calim}

The data were calibrated, imaged and self-calibrated using standard
techniques in the {\sc aips} package.  3C\,286 was used as the primary
amplitude calibrator if it was observable at sufficiently high
elevation during an observation, otherwise 3C\,48 was used.  We discuss the
flux-density scale in Section~\ref{scale}, below.  Our observations
were designed more for accurate imaging and polarimetry than for
flux-density measurement, and we did not determine gain-elevation or
opacity corrections independently. The transfer of the flux-density
scale was always done with secondary and primary calibrators observed
at similar elevations, however, so we do not expect large errors from
these effects over our frequency range.

All images were made using data from multiple configurations of
the VLA.  We did not attempt to match the spatial-frequency coverage
of the datasets in detail, but rather to ensure that the relevant
range of spatial scales was covered fully.  All of the visibility
datasets for the individual frequencies were imaged with scaled
maximum baselines, Gaussian tapers and data weighting to give the same
resolution and were deconvolved and restored with the same Gaussian
beam. We used three different deconvolution algorithms (see the
references in Table~\ref{tab:images}), depending on the complexity of
the source structure: conventional (single-resolution) {\sc clean};
multi-resolution {\sc clean} \citep{MSC} and maximum entropy
\citep[used as described by \citealt{LP91}]{CE}.  Data were usually
taken in two frequency channels simultaneously\footnote{The exceptions
  are early observations or those taken during phased-array VLBI
  runs.}. At frequencies above 4.5\,GHz, the two channels were imaged
together if they were contiguous.  The channels in the 1.3 -- 1.7\,GHz
band (typically more widely spaced in frequency) were imaged
separately. In a few of these cases where the centre frequencies of
two channels were less than 100\,MHz apart, the images were averaged
after deconvolution.

The registration of the images was assured during the imaging process
by shifting the visibility data so that the unresolved central component
(or ``core'') was always aligned on the centre of a given pixel (this was 
in any case required for accurate deconvolution).

\subsubsection{Flux-density scale}
\label{scale}

The accuracy and repeatability of the relative flux-density scale
across our frequency range are clearly important for this study.  We
initially used the values of flux density and visibility models for
3C\,286 and 3C\,48 provided with the {\sc aips} software distribution.
\citet{PB} recently carried out a comprehensive investigation of the
VLA flux-density scale. They showed that the flux density of 3C\,286
was stable over the period of our observations, but that of 3C\,48
varied significantly.  They derived improved flux densities for
both calibrators on the absolute scale established by WMAP
\citep{WMAP}, extrapolated to lower frequencies using a
thermophysical model of Mars. They also incorporated flux-density
measurements from \citet{Baars} at frequencies $\leq$3\,GHz. We
rescaled all of the images used in the present paper, multiplying by
the ratio of the flux density of the primary calibrator given by
\citet{PB} to that assumed in the initial data
reduction. Specifically, we used the polynomial expressions in
Tables~10 and 11 of \cite{PB} for our observing frequencies,
interpolating linearly to the relevant epoch for 3C\,48. Data taken
over periods of a few months to several years were combined to derive
the final images, so the variability of 3C\,48 between individual
observations was a potential concern.  For the relevant combinations,
its maximum variability amplitude was $<$1\%, however, so we 
ignored this complication.

The average effect of the recalibration determined by \cite{PB} is to
increase the measured flux densities at 1.3 -- 1.7\,GHz by
$\approx$2\% and to decrease those at 4.5 -- 5.0\,GHz and 8.3 --
8.5\,GHz by similar amounts.  The resulting change in spectral index
is between +0.03 and +0.04.  Sources are differently affected by the
variability of 3C\,48 and (very slightly) by inconsistencies in the
flux densities assumed for initial calibration.  The individual
correction factors are listed in Table~\ref{tab:images}.

\subsection{Lobe subtraction}
\label{sub}

Seven of the sources in our sample have lobes surrounding their jets
on scales of interest to this work (NGC\,193, 0206+35, 0326+39,
0755+37, 3C\,270, M\,84 and 3C\,296).  In six of these, the surface
brightness of the lobes is significant at arcsecond resolution (the
exception is 0326+39).  In order to isolate and measure the spectrum
of the jet emission, we need to remove the contribution of the lobe,
which typically has a steeper spectrum than the jets. To do this, we
used the method described by \citet{LB12}, assuming that the lobe
intensity varies smoothly and slowly across the jet. We first located
the edges of the flatter-spectrum jet emission on images of spectral
index and/or by using spectral tomography \citep{K-SR}.  We then
defined two background regions parallel to the jet axis and just
outside the maximum extent of the jet emission, smoothed the
background brightness distributions parallel to the jet axis with a
boxcar function and interpolated linearly between them under the jet.
The locations of the background regions and the widths of the
smoothing kernels are given in Table~\ref{tab:images}.

\subsection{Error model}
\label{erm}

Errors in measured brightness can be divided into three types: {\em random
  errors} which vary across individual images; {\em calibration
  errors} which multiply all pixels within a single image by the same
factor but are uncorrelated between images and {\em scale errors}
which affect all observations at given frequency in the same way.  

We took the random component to have additive and multiplicative
sub-components. For the additive part, we assumed that 
errors in total intensity have a Gaussian distribution with zero mean
and rms $\sigma_I$ in the image plane and that they are independent on
scales larger than the Gaussian restoring beam, which has an area of
$\pi f^2(2\ln 2)^{-1}$, where $f$ is the FWHM.  When computing
spectral-index images and determining blanking levels, we took
$\sigma_I$ to be the off-source noise level $\sigma_{\rm off}$
averaged over several areas well clear of any emission as given in
Table~\ref{tab:images}.  All of the sources in this study have bright,
unresolved cores, so the effect of residual calibration errors is to
increase the noise level at the jet bases.  When computing spectral
indices from flux densities integrated over different regions of the
jets we therefore followed the following recipe.
\begin{enumerate}
\item We used no data from points closer than $2f$ to the core.
\item We only included pixels in the summation if the random error on
  the spectral index at that point determined using the off-source
  noise levels $\sigma_{\rm off}$ was $\leq$0.03 (this automatically
  ensured that we integrated over the same points at all frequencies).
\item For points farther than $4f$ from the core, we took $\sigma_I = \sigma_{\rm off}$.
\item We adopted a different noise level $\sigma_{\rm core}$ for pixels
  located between $2f$ and $4f$ from the core. $\sigma_{\rm core}$,
  which is also given in Table~\ref{tab:images}, was estimated from
  rectangular regions displaced perpendicular to the jet axis on
  either side of the core.
\end{enumerate}
We also included a random error contribution which scales with surface
brightness, as might be expected for some types of calibration or
deconvolution artefacts.  Such effects are potentially quite
heterogeneous, ranging from offsets in the zero level due to missing
short-spacing coverage which affect an entire image, through coherent
ripples with various wavelengths to spotty structures caused by the
failures of the {\sc clean} algorithm.  For our images, they become
comparable with the additive component only after averaging over many
beamwidths.  On the basis of a comparison of images at the same or
neighbouring frequencies, we therefore crudely modelled this type of error as a
1\% rms multiplicative effect applied {\em to integrated flux
  densities only} (i.e.\ assumed to be coherent over the integration
region).  We also carefully reviewed all of the total intensity and
spectral-index images and identified only one source, M\,84, where
coherent calibration/deconvolution errors on scales larger than the
integration regions must be significant.  0206+35 shows artefacts at a
lower level, consistent with our error prescription. Both cases are
discussed in Appendix~\ref{notes}.

We next considered multiplicative calibration offsets.  Errors in the
gain calibration process may be partially correlated between datasets
for the same source, given that observations at different frequencies
were often interleaved. Nevertheless, all of the images were made from
at least two and often many more individual datasets, and we believe
that errors in the transfer of the flux-density scale from primary
calibrator to target are essentially independent. Additional errors
may be introduced during combination of data from different times
and/or array configurations. Although we always forced the mean
modulus of the complex gain correction to be unity during
self-calibration, the simultaneous adjustment of the relative
amplitude scale of the two datasets and the flux density of the
compact core, which may vary with time \citep{LB02a,LCBH06}, was not
always fully constrained at the higher frequencies, and additional
errors were introduced.  Our best estimate, derived from internal
consistency checks and in particular from a comparison of independent
measurements with the same array configuration(s) at similar or
identical frequencies, is that the appropriate rms calibration error
for an individual image is 3\% at all of our frequencies, or more
formally that the measured brightness is $kI$, where $\log k$ has a
Gaussian distribution with rms 0.0128.  We discuss the effects of
inaccuracies in this estimate at appropriate points below.  We treat
this type of error as uncorrelated between sources.

Finally, errors in the flux densities of the primary amplitude
calibrators introduce systematic scale errors for all observations at
a given frequency.  The absolute accuracy of the new VLA flux-density
scale is thought to be better than 2.5\% rms at 1.3 -- 1.7\,GHz and
1\% rms at our higher frequencies \citep{PB}. We are concerned with
the relative accuracy over our frequency range, which must be somewhat
better than this.

\subsection{Spectral-index fitting}
\label{spfit}

If images were available at only two frequencies $\nu_1$ and $\nu_2$,
we calculated the spectral indices directly as $\alpha^{\nu_2}_{\nu_1}
= \ln[I(\nu_1)/I(\nu_2)]/\ln(\nu_2/\nu_1)$ and determined the random
error $\sigma_{\rm r}$ from the rms errors in $I$ (following the
prescription in Section~\ref{erm}) by standard error propagation.  The
rms calibration error $\sigma_{\rm c}$ then just depends on the ratio
$\nu_2/\nu_1$, since we assume the same fractional error at all
frequencies:
\[
\sigma_{\rm c} = \frac{0.03 \times 2^{1/2}}{\ln(\nu_2/\nu_1)}
\]

Deviations from power-law spectra are of physical interest as
indicators of synchrotron losses and potentially as diagnostics of the
acceleration mechanism. For four sources (3C\,31, 0326+39, 1553+24 and
3C\,449), we have data in all of the 1.3 -- 1.7, 4.5 -- 5.0 and 8.3 --
8.5\,GHz bands and could potentially have addressed this issue. For
each source, we chose one frequency per band ($\nu_{\rm low}$,
$\nu_{\rm mid}$, $\nu_{\rm high}$) and plotted $\alpha^{\nu_{\rm
    high}}_{\nu_{\rm mid}}$ against $\alpha^{\nu_{\rm mid}}_{\nu_{\rm
    low}}$ (colour-colour plots; \citealt{KS93}) for the binned
profiles from Section~\ref{alphaprofiles} (see below).  For all four
sources, we found that the points were tightly clustered around single
locations in the colour-colour diagram, consistent with the effect of
a calibration error.  From our error model (Section~\ref{erm}), we predict
an rms for the spectral-index difference $\alpha^{\nu_{\rm
    high}}_{\nu_{\rm mid}} - \alpha^{\nu_{\rm mid}}_{\nu_{\rm low}}$
of 0.10; we found an rms of 0.12 and a mean of +0.05 for the four
sources, with one (3C\,31) showing spectral flattening with increasing
frequency.  We conclude that any apparent spectral curvature observed 
over our relatively narrow frequency range is entirely consistent with the
effects of calibration errors.  Fig.~13 of \citealt{3c31ls} shows integrated
spectra which reinforce this point.

Given our lack of evidence for real spectral curvature, we chose to fit
single power laws to spectra with three or more frequencies and
determined the spectral indices by least-squares fitting, using
weights derived from the random and systematic errors, summed in
quadrature.  We then took $\sigma_{\rm c}$ to be the error in the
limit of zero random noise and $\sigma_{\rm r} = (\sigma^2_{\rm fit} -
\sigma^2_{\rm c})^{1/2}$, where $\sigma_{\rm fit}$ is the error
estimate from the fit.

For consistency, all of the displayed spectral-index images (whether
determined from two frequencies or more) are blanked wherever
$\sigma_{\rm r} > 0.03$, assuming constant noise levels $\sigma_I =
\sigma_{\rm off}$ across the maps to calculate $\sigma_{\rm r}$
(i.e.\ without multiplicative random or systematic errors). This
criterion is also part of that used to select the points included in
the integrated flux densities (Section~\ref{erm}, above).

\section{Jet models}
\label{Models}

\begin{table}
\caption{Fiducial distances derived from jet models, projected on the
  plane of the sky, and details of jet detections at higher
  frequencies.  (1) name; (2) recollimation distance $z_0$;
  (3) start of high-emissivity region (flaring point), $z_{\rm
    e1}$; (4) end of high-emissivity region $z_{\rm e0}$;
  (5) start of deceleration, $z_{\rm v1}$; (6) end of
  deceleration $z_{\rm v0}$; (7) maximum length of detected X-ray
  or optical emission $z_{\rm OX}$ for the main jet; (8) notes
  and references for high-frequency detections.  All distances are in
  arcsec. Only the recollimation distance and the position of the
  flaring point are given for 3C\,449: these are the two parameters
  that can be determined directly from the brightness distribution
  without a model.\label{tab:fiducials}}
\begin{minipage}{8.5cm}
\begin{tabular}{lrrrrrrr}
\hline
&&&&&&&\\
Name        &\multicolumn{6}{c}{Distances / arcsec}&Note\\
            &$z_0$&$z_{\rm e1}$&$z_{\rm e0}$&$z_{\rm v1}$&$z_{\rm v0}$&$z_{\rm OX}$&\\
&&&&&&&\\
\hline
&&&&&&&\\
NGC\,193    &  22.2 &   3.3  &  7.0&   0.0&  13.9&    5 & 1\\
NGC\,315    &  79.5 &   5.8  &17.7 & 15.1 & 36.8 &    30& 2\\
3C\,31      &   8.5 &   2.8  & 8.5 &  5.4 &  9.5 &    8 & 3\\
0206+35     &   4.5 &   0.7  & 1.8 &  1.5 &  3.5 &    2 & 4\\
0326+39     &  18.2 &   1.8  & 5.1 &  3.0 & 12.4 &      &  \\
0755+37     &   9.4 &   1.0  & 9.0 &  2.5 & 18.7 &    4 & 5\\      
3C\,270     &  34.0 &   9.0  &25.3 & 13.0 & 28.7 &    32& 6\\
M\,84       &  17.0 &   2.8  & 6.8 &  2.7 & 13.2 &    4 & 7\\
3C\,296     &  27.5 &   3.0  &13.0 &  8.9 &  9.8 &    10& 8\\
1553+24     &   6.5 &   0.5  & 1.7 &  3.7 &  4.2 &    1 & 9\\
3C\,449     &  11.0 &   4.5  &     &      &      &      & \\
&&&&&&&\\
\hline
\end{tabular}
\noindent References. Unless noted explicitly, these describe X-ray observations.\\
1 \citet{Kharb12}\\
2 \citet{Worrall07}\\
3 \citet{H02}; optical and mid-IR emission have also been detected from the main jet 
\citep{Lanz,Croston03}\\
4 \citet*{WBH01}\\
5 \citet{WBH01}; the inner 2.6\,arcsec of the main jet has been detected at optical wavelengths 
\citep{Parma03}\\ 
6 \citet{Worrall10}; the inner 20\,arcsec of the counter-jet was detected in the same observation.\\
7 \citet{Harris02}\\
8 \citet{H05}; optical emission from the main jet is also reported in this reference.\\
9 \citet{Parma03}; this is a detection of optical emission.\\
\end{minipage}
\end{table}

Our models assume that the jets are symmetrical, axisymmetric and
relativistic \citep{LB02a,CL,CLBC,LCBH06,LB12}.  Relativistic
aberration not only causes the approaching (main) jet to appear
brighter than the receding counter-jet, but also leads to differences
in observed linear polarization, as the angles to the line of sight in
the rest frame of the emission are different on the two sides.
Simultaneous fitting of Stokes $I$, $Q$ and $U$ then allows us to
break the degeneracy between velocity and inclination.  The basic
steps in our method are:
\begin{enumerate}
\item construct a parameterized model of geometry, velocity, emissivity and field-ordering in the rest frame;
\item calculate the observed-frame emission in $I$, $Q$ and $U$;
\item integrate along the line of sight and convolve with the observing beam;
\item evaluate $\chi^2$ between model and observed images and
\item optimize the model parameters.
\end{enumerate}

\begin{figure}
\begin{center}
\includegraphics[width=8.5cm]{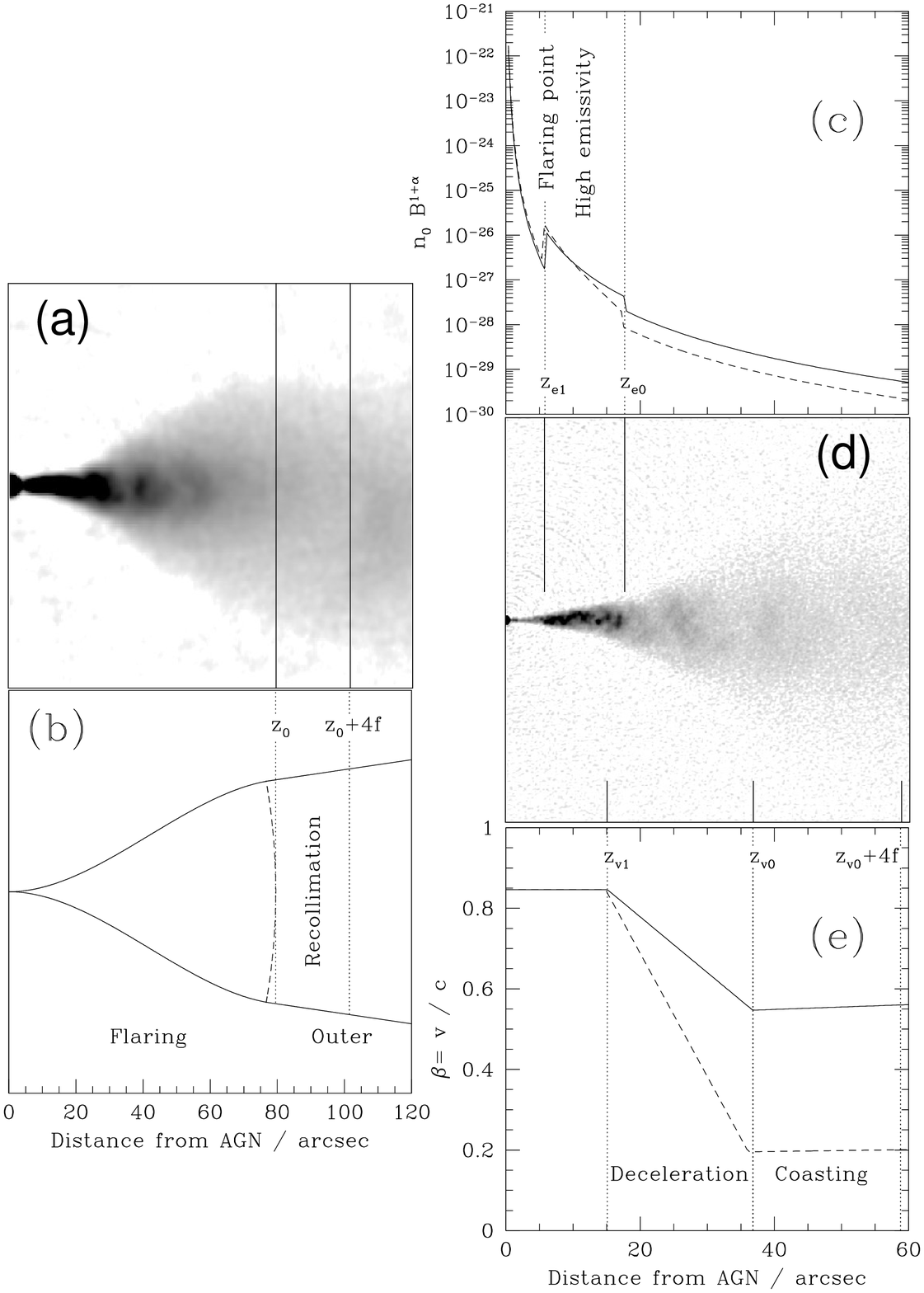}
\caption{Jet models and fiducial regions for spectral indices. (a) The
  main jet of NGC\,315 at 2.35-arcsec FWHM resolution, indicating the
  area over which $\alpha_{\rm recoll}$ was evaluated. (b) Sketch
  showing the flaring and outer regions of the jets in NGC\,315 to the
  same scale as panel (a), but with no bend in the jet. (c)
  Longitudinal profiles of rest-frame emissivity derived from our
  model of NGC\,315 and plotted against distance from the AGN
  projected on the plane of the sky. The high-emissivity region is
  marked. (d) Grey-scale of the main jet at a resolution of
  0.4\,arcsec FWHM, showing the high-emissivity, deceleration and coasting
  regions. (e) Longitudinal profiles of velocity derived from our
  model of NGC\,315, with the deceleration and coasting regions
  marked.  In panels (c) and (e), the full and dashed lines apply to
  the axis and edge of the jet, respectively.
\label{fig:sketch}}
\end{center}
\end{figure}

In order to make physically meaningful comparisons between spectral
indices in different jets, we defined integration regions based on
three aspects of our model fits: geometry, proper emissivity and
velocity.  In Fig.~\ref{fig:sketch}, we illustrate these regions using
the observed main-jet brightness distribution and model fits for
NGC\,315 (Laing \& Bridle, in preparation). The relevant aspects of
our model fits are as follows.
\begin{description}
\item [{\bf Geometry}] We divide a jet into two regions based on the
  geometry of its outer isophotes: a {\em flaring region}, where the
  flow first expands with increasing opening angle and then
  recollimates and a conical {\em outer region}
  (Figs~\ref{fig:sketch}a and b).  We define $z_0$ to be the distance
  between the AGN and the boundary between the regions, measured
  along the axis and projected on the plane of the sky.
\item [{\bf Proper emissivity}] We model the quantity $n_0
  B^{1+\alpha}$, where $n_0$ is the normalizing constant in the
  electron energy distribution and $B$ is the rms total magnetic
  field, both measured in the rest frame.  $n_0$ and $B$ cannot be
  separated using observations of synchrotron emission without
  additional assumptions, for example flux-freezing \citep{LB04} or
  equipartition between particle and field energy. We refer to $n_0
  B^{1+\alpha}$, loosely, as ``the emissivity''; in fact it is
  multiplied by a constant and by factors dependent on field geometry
  (which we ignore in the present discussion) to give the true
  emissivities in $I$, $Q$ and $U$ \citep{L02}.  The emissivity
  profile along the jet axis is divided into 3 or 4 sections, in each
  of which the emissivity has a power-law dependence on distance from
  the nucleus.  All of the jets have a well-defined {\em flaring
    point} marked by an abrupt increase in brightness. We model this
  as a step in emissivity and a change in power-law slope.  The
  flaring point marks the start of the {\em high-emissivity region},
  which is characterized by bright, knotty, non-axisymmetric
  substructure and which is usually obvious by eye in the brightness
  distribution of the main jet. The end of this region is again marked
  by a step in emissivity and/or a change (usually flattening) of
  slope.  These features are illustrated for NGC\,315 in
  Figs~\ref{fig:sketch}(c) and (d). The positions of the start and end
  of the high-emissivity region from the AGN, measured from the
  nucleus along the projection of the jet axis on the sky, are $z_{\rm
    e1}$ and $z_{\rm e0}$, respectively.
\item [{\bf Velocity}] The on-axis velocity is modelled as initially
  constant, decreasing linearly with distance in the {\em deceleration
    region} and then varying, again linearly but much more slowly, at large
  distances (we allow either acceleration or deceleration in the
  fits). Deceleration usually begins within the high-emissivity region
  and ends before recollimation.  The start and end of the
  deceleration region, again measured from the AGN along the
  projected jet axis, are $z_{\rm v1}$ and $z_{\rm v0}$, respectively.
\end{description}
A full tabulation of fitted model parameters will be given by Laing \&
Bridle (in preparation). In the present paper, we use only the
fiducial distances and (in Section~\ref{alpha-vel}), the velocity
profiles.  We list the angular distances $z_0$, $z_{\rm e1}$, $z_{\rm
  e0}$, $z_{\rm v1}$ and $z_{\rm v0}$ in
Table~\ref{tab:fiducials}. These quantities define our integration
regions. Other aspects of our models, such as the shapes of the
streamlines and boundary surfaces between regions, the transverse
variations of velocity and emissivity and the distribution of
magnetic-field component ratios are less relevant to the present
discussion.

We defined four integrated spectral indices, as follows.
\begin{description}
\item [{\bf High-emissivity, $\alpha_{\rm he}$}] We integrated over the
  part of the high-emissivity region that is farther than $2f$ from
  the core (Figs~\ref{fig:sketch}c and d), stopping the integration at
  the end of the deceleration region if the high-emissivity region
  extends beyond it.
\item [{\bf Deceleration, $\alpha_{\rm decel}$}] We summed over the
  deceleration region, stopping at recollimation if this occurs before
  the end of deceleration.
\item [{\bf Coasting, $\alpha_{\rm coast}$}] We integrated 
  from the end of the deceleration region for $4f$ or until
  recollimation, whichever is closer.
\item [{\bf Recollimation, $\alpha_{\rm recoll}$}] The integration started  
  at recollimation and extended for $4f$ into the outer region.
\end{description} 
The high-emissivity and deceleration regions usually overlap, but the
definitions were designed to ensure that the coasting and
recollimation regions are disjoint with each other and with the
remaining two regions.  The resolution of our observations is too low
to measure reliably the spectral index of emission any closer to the
AGN except in the main jet of 3C\,270 (Appendix~\ref{notes}).

We integrated the flux densities over all pixels satisfying the
criteria of Section~\ref{erm}, requiring in addition that the total
number of unblanked pixels in the region, $N \geq 2N_{\rm beam}$,
where $N_{\rm beam} = \pi (f/\Delta x)^2(2\ln 2)^{-1}$ is the beam
area in pixels of size $\Delta x$. We then calculated spectral indices
and errors as in Section~\ref{spfit}.

\begin{figure*}
\begin{center}
\includegraphics[width=15cm]{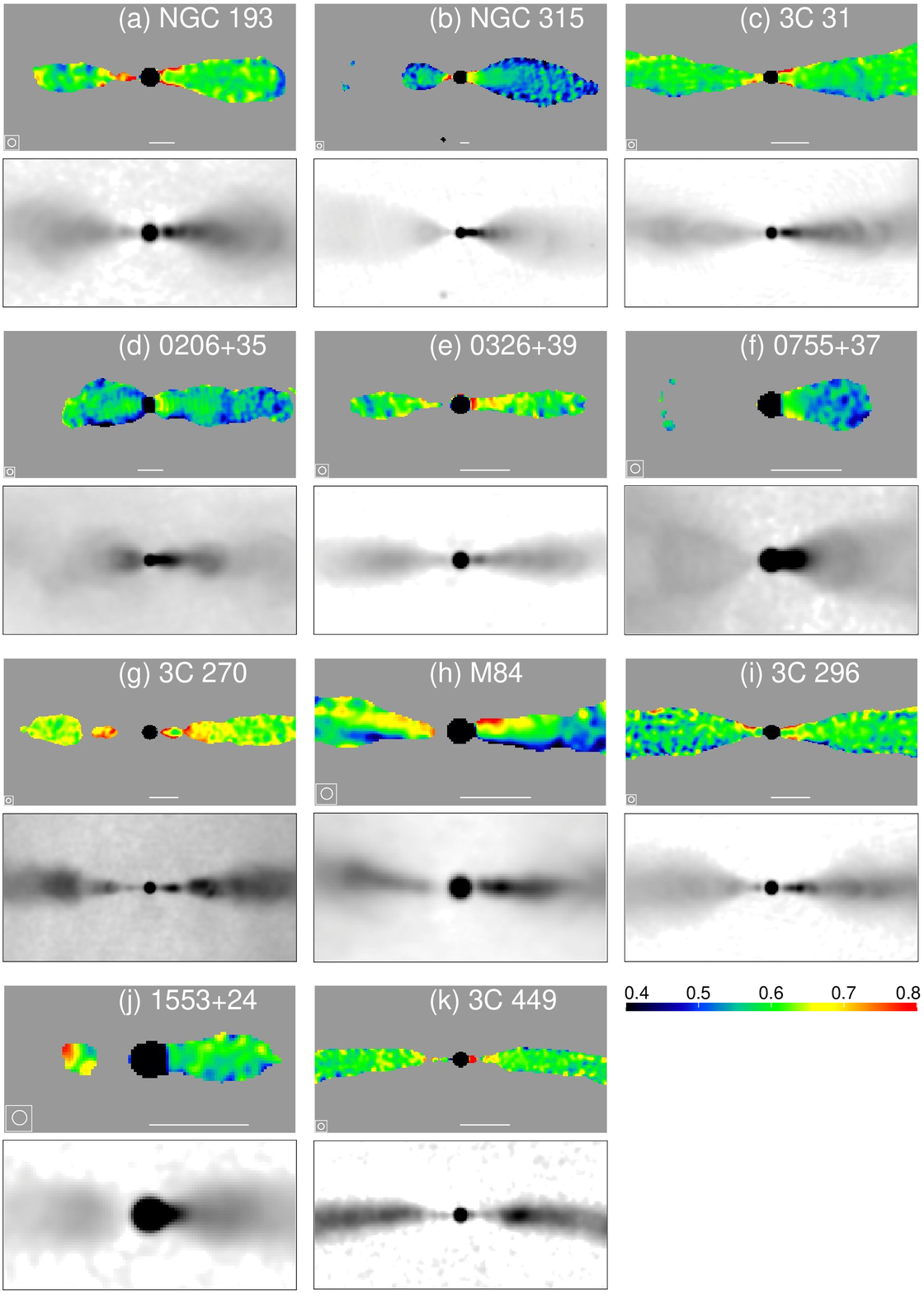}
\caption{False-colour images of spectral index, $\alpha$ (upper
  panels) and grey-scales of total intensity (lower panels) at the
  same resolution.  $\alpha$ is in the range 0.4 -- 0.8 (as indicated
  by the labelled wedge) and is plotted wherever $\sigma_{\rm r} <
  0.03$.  The images have been rotated by the angles given in
  Table~\ref{tab:images}, so that the approaching jets are always on
  the right. On the spectral-index plots, the horizontal lines
  represent distances of 10\,arcsec and the restoring beams are shown
  in the bottom-left-hand corners. Lobe emission has not been
  subtracted from the $I$ images.
\label{fig:images}}
\end{center}
\end{figure*} 

\begin{figure*}
\begin{center}
\includegraphics[width=15cm]{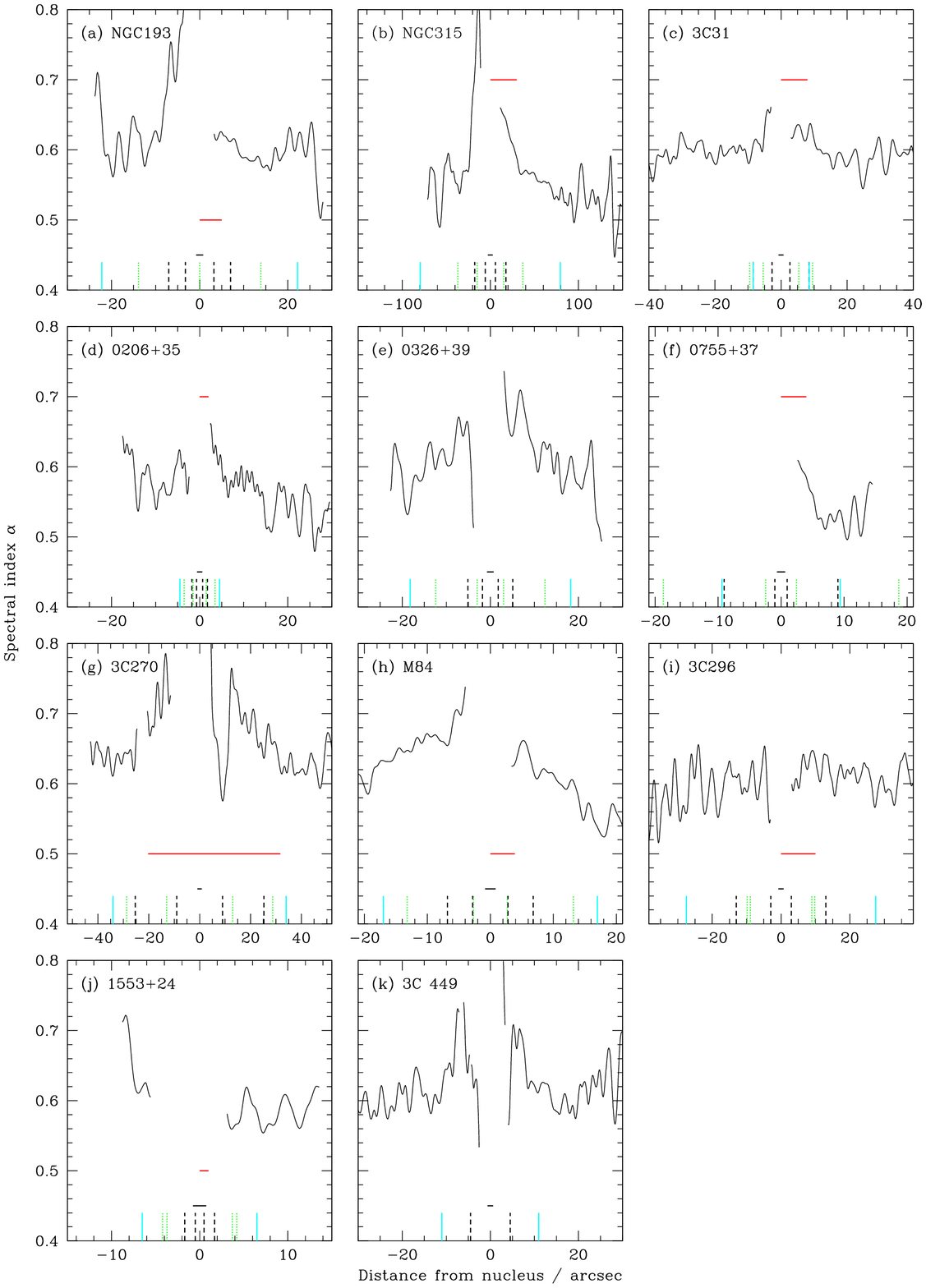}
\caption{Profiles of spectral index, $\alpha$, along the jets. The
  brighter (approaching) jet is always on the right. The profiles are
  derived from the images shown in Fig.~\ref{fig:images}, along
  horizontal lines passing through the nuclei (a different position
  angle was used for the counter-jet of M\,84, which is misaligned
  with the main jet).  Profiles are only plotted where $\sigma_{\rm r}
  \leq 0.03$ and the distance from the core exceeds 2 $\times$
  FWHM. The beamwidths (FWHM) are indicated by the horizontal lines
  centred on distance 0 and $\alpha = 0.45$. The vertical dashed black
  lines represent the extent of the high-emissivity region and the
  dotted vertical green lines the start and end of rapid
  deceleration. The vertical full cyan lines show where the jets
  recollimate. The horizontal red lines plotted at $\alpha = 0.5$ or
  $\alpha = 0.7$ show the maximum extent of any detected soft X-ray or
  optical emission from the jets (Table~\ref{tab:fiducials}).
\label{fig:profiles}}
\end{center}
\end{figure*} 
\begin{figure*}
\begin{center}

\includegraphics[width=15cm]{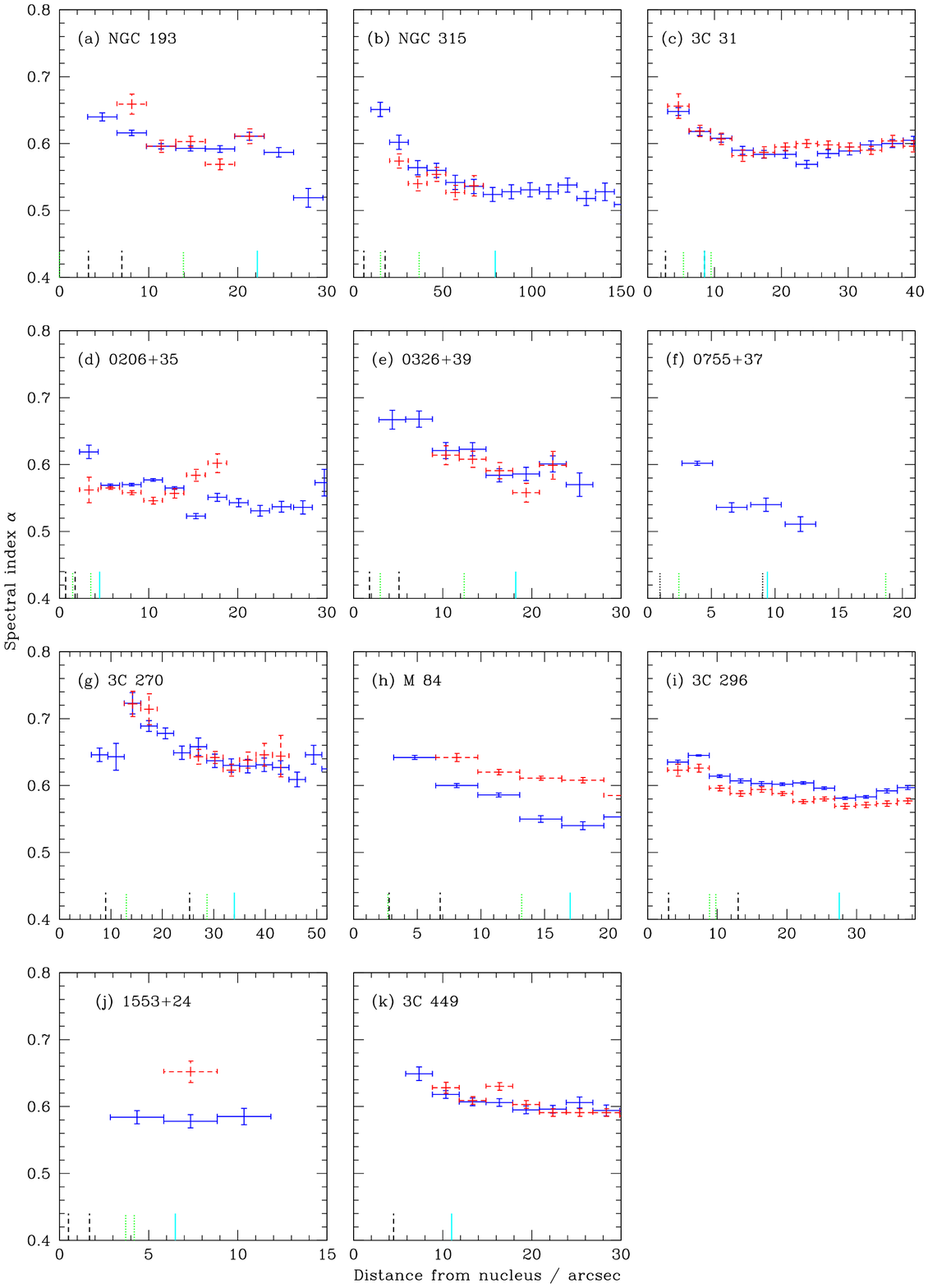}
\caption{Profiles of integrated spectral index, $\alpha$, along the
  jets.  The spectral indices were calculated from total intensities
  integrated over points with $\sigma_{\rm r} \leq 0.03$ in boxes of
  length 2FWHM along the jet axis (horizontal in
  Fig.~\ref{fig:images}, starting at 2FWHM from the core. Data are
  only plotted if there are more than two independent beam areas above
  the blanking level in the box. Blue full and red dashed symbols
  represent main and counter-jets, respectively. The fiducial
  distances are indicated as in Fig.~\ref{fig:profiles}.
\label{fig:binnedprofiles}}
\end{center}
\end{figure*} 

\section{Spectral-index images and profiles}
\label{alphas}

\subsection{Images}
\label{alphaimages}

In Fig.~\ref{fig:images} we show false-colour images of spectral index
over the range $0.4 \leq \alpha \leq 0.8$ for all of the sources,
together with grey-scales of total intensity at the same
resolution. The images have been rotated by the angles given in
Table~\ref{tab:images} so as to have the main jets on the right.

Except for the unresolved cores, which are partially optically thick with $\alpha
\la 0.3$, the emission typically has $0.7 \ga \alpha \ga
0.5$. Significantly higher or lower spectral indices are only found
close to the edges of the jets (where small errors in deconvolution, lobe subtraction or
image registration can cause large changes in $\alpha$) or in regions
of low signal-to-noise ratio.  The mean spectral indices of the jets
vary over a range of $\approx$0.1, with NGC\,315 showing the flattest
spectrum and 3C\,270 the steepest.  The spectral indices of main and
counter-jets are usually very similar, with M\,84 and 1553+24 the only
conspicuous exceptions.

A slight, but significant tendency for the spectral index to decrease
away from the AGN is apparent from the images in
Fig.~\ref{fig:images}.  The effect is subtle ($\Delta\alpha \la 0.1$)
but consistent.

\begin{table*}
\caption{Integrated spectral indices for high-emissivity, deceleration, 
  coasting and recollimation regions. (1) source name; (2) 
  rms calibration error for spectral indices $\sigma_{\rm c}$. The
  remaining columns give the mean spectral indices, $\langle\alpha\rangle$ and their
  estimated random errors for the four fiducial regions in each
  jet. Values are not given if one or more of the regions: has too
  little data of adequate signal-to-noise (NGC\,315; 3C\,31, 0755+37), is too
  close to the core (0206+35 and 1553+24) or is undefined (3C\,449).
\label{tab:numbers}}
\begin{tabular}{llllllllllllllllll}
\hline
&&&&&&&&&&&&&&&&&\\ 
Source &$\sigma_{\rm c}$&\multicolumn{4}{c}{High-emissivity}&\multicolumn{4}{c}{Deceleration}&
\multicolumn{4}{c}{Coasting}&\multicolumn{4}{c}{Recollimation}\\
       &                &\multicolumn{2}{c}{jet}&\multicolumn{2}{c}{cj}&\multicolumn{2}{c}{jet}&\multicolumn{2}{c}{cj}&\multicolumn{2}{c}{jet}&\multicolumn{2}{c}{cj}&\multicolumn{2}{c}{jet}&\multicolumn{2}{c}{cj}\\
       & & $\langle\alpha\rangle$ & $\pm$ & $\langle\alpha\rangle$ & $\pm$& $\langle\alpha\rangle$ & $\pm$& $\langle\alpha\rangle$ & $\pm$& $\langle\alpha\rangle$ & $\pm$& $\langle\alpha\rangle$ & $\pm$& $\langle\alpha\rangle$ & $\pm$& $\langle\alpha\rangle$ & $\pm$\\
&&&&&&&&&&&&&&&&&\\
\hline
&&&&&&&&&&&&&&&&&\\
NGC  193&0.036&0.64&0.01&0.75&0.05&0.62&0.01&0.63&0.02&0.59&0.01&0.59&0.02&0.58&0.02&0.65&0.03\\
NGC  315&0.027&0.65&0.01&0.76&0.04&0.60&0.01&0.57&0.01&0.55&0.01&0.55&0.01&0.53&0.01&    &    \\
3C  31  &0.017&0.64&0.01&0.64&0.01&0.63&0.01&0.63&0.01&    &    &    &    &0.60&0.01&0.60&0.01\\
0206+35 &0.036&    &    &    &    &0.63&0.02&0.56&0.05&0.59&0.03&0.57&0.04&0.56&0.01&0.56&0.01\\
0326+39 &0.023&0.67&0.01&0.60&0.09&0.65&0.01&0.62&0.01&0.60&0.01&0.60&0.01&0.59&0.01&0.57&0.01\\
0755+37 &0.036&0.58&0.01&    &    &0.57&0.01&    &    &    &    &    &    &    &    &    &    \\
3C  270 &0.036&0.67&0.02&0.71&0.03&0.68&0.02&0.67&0.02&0.64&0.02&0.63&0.02&0.63&0.02&0.64&0.02\\
M  84   &0.036&0.64&0.01&0.67&0.02&0.61&0.01&0.63&0.01&0.55&0.01&0.61&0.01&0.55&0.01&0.59&0.01\\
3C  296 &0.034&0.64&0.01&0.62&0.01&0.63&0.01&0.62&0.01&0.60&0.01&0.59&0.01&0.58&0.01&0.57&0.01\\
1553+24 &0.023&    &    &    &    &0.57&0.02&    &    &0.60&0.01&0.61&0.04&0.58&0.01&0.67&0.02\\
3C  449 &0.016&0.68&0.01&0.68&0.03&    &    &    &    &    &    &    &    &0.61&0.01&0.61&0.01\\
&&&&&&&&&&&&&&&&&\\
\hline
\end{tabular}
\end{table*}

\subsection{Longitudinal profiles}
\label{alphaprofiles}

In order to investigate the spectral gradients in more detail, we
have plotted profiles of spectral index along the jets in two
representations: as unaveraged slices along the jet ridge-lines to
show the level of variation on small as well as large scales
(Fig.~\ref{fig:profiles}) and as binned averages to reduce random
noise and to compare values in the main and counter-jets
(Fig.~\ref{fig:binnedprofiles}).  The unbinned profiles are plotted
for distances $>$2$f$ from the AGN wherever the rms random error
on the spectral index determined from the off-source noise
$\sigma_{\rm off}$ (Table~\ref{tab:images}) is $\sigma_{\rm r} \leq
0.03$.  The binned profiles follow the prescription given in
Section~\ref{erm}.

The clear tendency for the spectra to flatten with distance from the
AGN is confirmed, and the binned profiles
(Fig.~\ref{fig:binnedprofiles}) show in addition that the jet and
counter-jet spectral indices at the same distance from the AGN are
usually very similar.  Of the two exceptions noted earlier, M\,84
(Fig.~\ref{fig:binnedprofiles}h) shows an offset between the spectral
indices of its jets which results from an instrumental effect
(Appendix~\ref{notes}) and 1553+24 (Fig.~\ref{fig:binnedprofiles}j)
has a weak counter-jet with measurable spectral index only over a
limited range of distance.

The fiducial distances from our jet models (projected on the sky) are
plotted in Figs~\ref{fig:profiles} and \ref{fig:binnedprofiles}. {\em
  Most of the spectral flattening with distance from the AGN occurs
  before the jets recollimate, and appears to be associated with rapid
  deceleration.} The clearest examples are the well-resolved and
bright sources NGC\,315, 3C\,31, 3C\,270 and 3C\,296
(Figs~\ref{fig:binnedprofiles}b, c, g and i).

High-frequency ($>$10$^{14}$\,Hz) emission is seen in many of the main
jets and in the counter-jet of  3C\,270.  Although the data are
heterogeneous, with a mixture of optical and X-ray detections and
various exposure times, the comparison in Fig.~\ref{fig:profiles}
shows that the detected high-frequency emission typically extends
roughly as far as the end of the high-emissivity region.
 
\subsection{Spectral indices at fiducial locations}
\label{averages}

\begin{figure}
\begin{center}
\includegraphics[width=4.25cm]{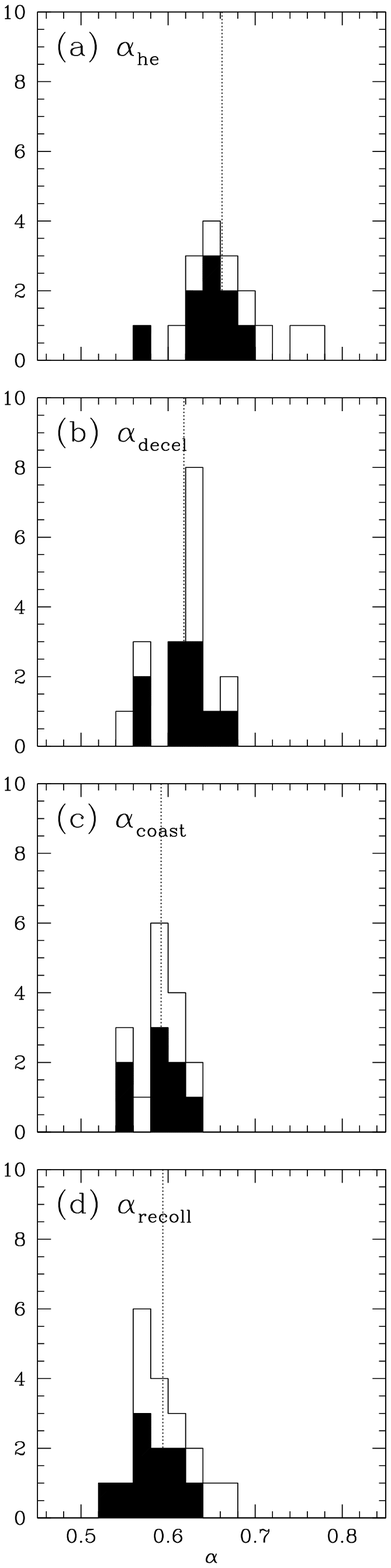}
\caption{Histograms of spectral indices at different locations, as
  given in Table~\ref{tab:numbers}.  (a) high-emissivity; (b)
  deceleration; (c) coasting; (d) recollimation.  The shaded
  areas represent the main jets. The vertical dotted lines indicate
  the unweighted means of the distributions, from
  Table~\ref{tab:means}.
\label{fig:alphahist}}
\end{center}
\end{figure} 

\begin{figure}
\begin{center}
\includegraphics[width=8.5cm]{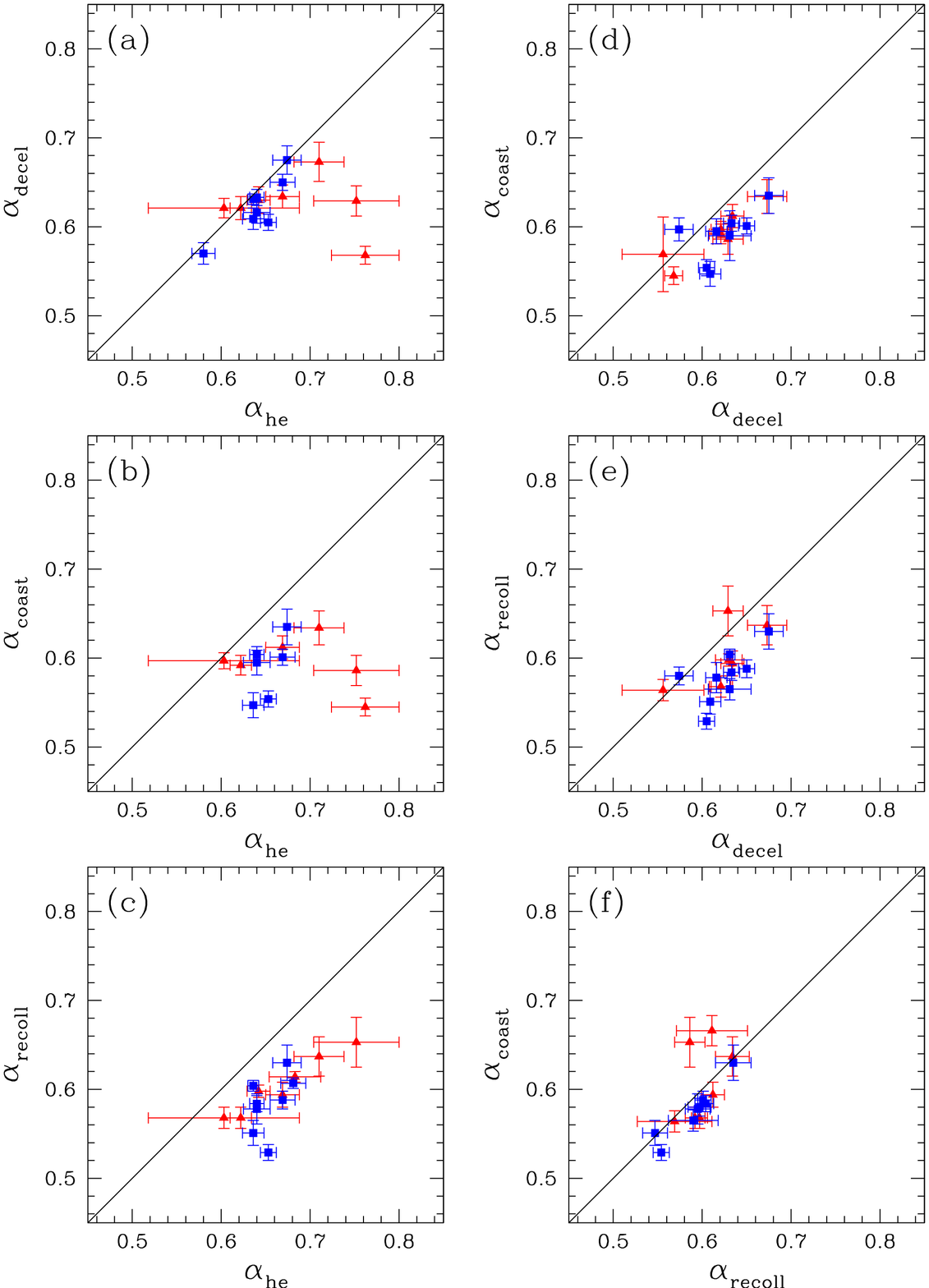}
\caption{Plots comparing spectral indices at different locations in
  the same jet.  (a) High-emissivity and deceleration. (b)
  High-emissivity and coasting. (c) High-emissivity and
  recollimation. (d) Deceleration and coasting. (e)
  Deceleration and recollimation. (f) coasting and
  recollimation. Blue squares: main jets; red triangles: counter-jets. The
  error bars correspond to random errors $\sigma_{\rm r}$ only: calibration
  errors would be expected to move the point parallel to the line of
  equality, which is plotted for reference.
\label{fig:alpha_alpha}}
\end{center}
\end{figure} 

\begin{figure}
\begin{center}
\includegraphics[width=4.25cm]{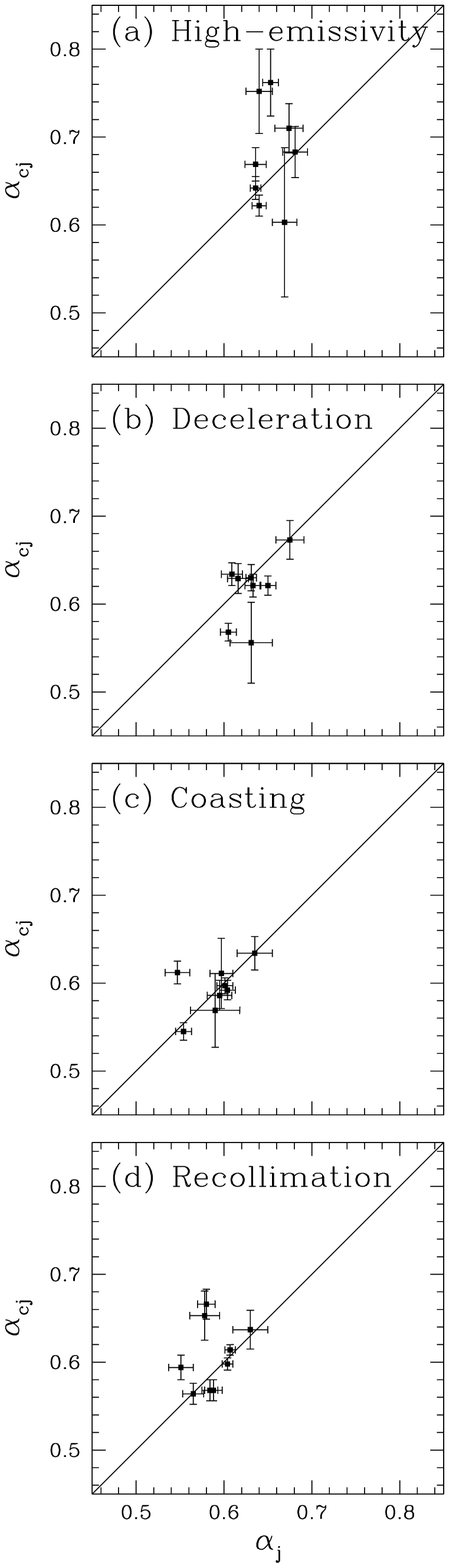}
\caption{Comparison between the main and counter-jet spectral indices,
  $\alpha_{\rm j}$ and $\alpha_{\rm cj}$ for the four fiducial
  regions. (a) high-emissivity; (b) deceleration; (c) coasting; (d)
  recollimation. The error-bars represent the random errors,
  $\sigma_{\rm r}$.
\label{fig:mj_cj}}
\end{center}
\end{figure} 

\begin{table}
\caption{Mean and rms spectral indices for the four fiducial regions,
  for main and counter-jets separately and for the
  combination. (1) jet class; (2) location; (3) number of
  jets; (4) unweighted mean spectral index; (5) measured
  spectral index rms; (6) rms predicted from the combination of
  random and calibration errors; (7) standard error on the unweighted
  mean; (8) mean weighted by random and calibration
  errors.\label{tab:means}}.
\begin{tabular}{llrlllll}
\hline
&&&&&&&\\                                                   
Jet & Region&$N$&\multicolumn{4}{c}{Unweighted}&Wtd\\
    &       &   &$\langle\alpha\rangle$&$\sigma_{\rm obs}$&$\sigma_{\rm pred}$&$\pm$&$\langle\alpha\rangle$\\
&&&&&&&\\                                                   
\hline
&&&&&&&\\                                                   
Main    & high-         & 9& 0.645& 0.028&0.033&0.009&0.650\\
CJ      & emiss-        & 8& 0.680& 0.054&0.050&0.019&0.668\\
Both    & ivity         &17& 0.662& 0.046&0.042&0.011&0.656\\ 
&&&&&&&\\                                             
Main    & decel-        &10& 0.619& 0.031&0.034&0.010&0.621\\ 
CJ      & eration       & 8& 0.617& 0.035&0.038&0.012&0.618\\ 
Both    &               &18& 0.618& 0.033&0.036&0.008&0.620\\ 
&&&&&&&\\                                             
Main    & coast-        & 8& 0.590& 0.026&0.036&0.009&0.590\\ 
CJ      & ing           & 8& 0.593& 0.026&0.040&0.009&0.590\\ 
Both    &               &16& 0.592& 0.026&0.038&0.007&0.590\\ 
&&&&&&&\\                                                                                                         
Main    & recoll-       &10& 0.582& 0.027&0.032&0.009&0.589\\ 
CJ      & imation       & 9& 0.607& 0.036&0.034&0.012&0.605\\ 
Both    &               &19& 0.594& 0.034&0.033&0.008&0.596\\ 
&&&&&&&\\                                                   
\hline
\end{tabular}
\end{table}

\begin{table}
\caption{Differences between spectral indices integrated over the
  standard regions. (1) spectral-index combination; (2) number of
  measurements (main and counter-jets); (3) mean spectral-index
  difference; (4) rms; (5) standard error on the mean;
  (6) significance level of correlation between spectral indices
  at the two locations, from the Spearman rank
  test.\label{tab:alphadiffs}}
\begin{tabular}{lrrrrr}
\hline
&&&&&\\
Location & N & $\langle\Delta\alpha\rangle$ & rms & $\pm$ &sig\\
&&&&&\\
\hline
&&&&&\\
$\alpha_{\rm he}-\alpha_{\rm decel}$ & 15 & 0.035 & 0.053 & 0.014       & 77.0\\
$\alpha_{\rm he}-\alpha_{\rm coast}$ & 12 & 0.077 & 0.058 & 0.017   & 18.0\\
$\alpha_{\rm he}-\alpha_{\rm recoll}$ & 15 & 0.067 & 0.024 & 0.006      &   99.9\\
$\alpha_{\rm decel}-\alpha_{\rm coast}$ & 15 & 0.029 & 0.022 & 0.006 &$>$99.9\\
$\alpha_{\rm decel}-\alpha_{\rm recoll}$ & 16 & 0.037 & 0.027 & 0.007   & 99.7 \\
$\alpha_{\rm coast}-\alpha_{\rm recoll}$ & 15 & 0.005 & 0.028 & 0.007&99.9\\
&&&&&\\
\hline
\end{tabular}
\end{table}

\begin{table}
\caption{Mean differences between the spectral indices of the main and
  counter-jets at the four  fiducial locations. (1) location; (2) mean $\langle \alpha_{\rm cj} - \alpha_{\rm j}\rangle$; (3) standard error on the mean; (4) number of sources; (5) significance level of correlation between $\alpha_{\rm cj}$ and $ \alpha_{\rm j}$, according to the Spearman rank test.\label{tab:mj_cj}}
\begin{tabular}{lrrrr}
\hline
&&&&\\
Location &$\langle \alpha_{\rm cj} - \alpha_{\rm j}\rangle$&$\pm$& N & sig\\
&&&&\\
\hline
&&&&\\
High-emissivity & 0.027 & 0.020 & 8 & 35.3\\
Deceleration &-0.015 & 0.010 & 8 & 43.5\\
Coasting & 0.003 & 0.009 & 8 &68.0\\
Recollimation & 0.019 & 0.012 & 9 &47.1\\
&&&&\\
\hline
\end{tabular}
\end{table} 

Individual integrated spectral indices over the four fiducial areas 
are given in Table~\ref{tab:numbers}, along with rms random errors
$\sigma_{\rm r}$ estimated using the prescription in
Section~\ref{erm}.  The rms calibration errors $\sigma_{\rm c}$, which
are common to all of the measurements for an individual jet, are also
tabulated. $\sigma_{\rm r} \approx 0.01 - 0.02$ in regions of high
signal-to-noise, with significantly larger values primarily in those
counter-jets whose high-emissivity regions are both faint and close to
a bright core.

Histograms of the distributions are plotted in
Fig.~\ref{fig:alphahist}.  These display the progressive decrease of 
spectral index from high-emissivity through deceleration to
coasting regions. {\em There is no evidence of further systematic spectral 
flattening after recollimation:} the distributions for the coasting and
recollimation regions are statistically the same (the significance
level of the difference is only 28\% according to the
Kolmogorov-Smirnov test), consistent with very similar spectral-index
and error distributions.  The histogram for the high-emissivity region
(Fig.~\ref{fig:alphahist}a) is slightly broader than the other three;
this is expected from the larger random errors in the counter-jet
spectral indices, and the distribution for the main jets alone is
narrower.

Statistics of the distributions for the main and counter-jets
separately and for the two combined are listed in
Table~\ref{tab:means}.  Here, and wherever we quote the error on a
mean spectral index, we give the standard error derived from the
dispersion in the measurements. The unweighted means are very close to
those weighted by the combined (random and calibration) errors.  The
unweighted mean spectral index for the high-emissivity region,
$\langle \alpha_{\rm he} \rangle = 0.662\pm 0.011$, is significantly
steeper than those for either the coasting or the recollimation
regions ($\langle \alpha_{\rm coast} \rangle = 0.592 \pm 0.007$ and
$\langle \alpha_{\rm recoll} \rangle = 0.594 \pm 0.008$,
respectively). The mean for the deceleration region, $\langle
\alpha_{\rm decel} \rangle = 0.618\pm 0.008$, is in the middle of this
range.  While both the random and calibration errors contribute to the
observed dispersion, any systematic error in the flux-density scale
does not.  We expect the rms scale error in spectral index to be
$<$0.02 (the value corresponding to independent scale errors at 1.4 and
4.9\,GHz as given by \citealt{PB}).

More accurate measures of the {\em change} in spectral index with
distance are given by the differences for individual jets, since any
scale and calibration errors then cancel out.  We plot the combinations of spectral
indices at the four standard locations in
Fig.~\ref{fig:alpha_alpha}. All of the combinations show spectral
flattening with increasing distance except for the plot of
$\alpha_{\rm recoll}$ against $\alpha_{\rm coast}$
(Fig.~\ref{fig:alpha_alpha}f). In particular, {\em all} of the jets
have spectra which flatten between the high-emissivity region and both
the coasting and recollimation regions
(Figs~\ref{fig:alpha_alpha}b and c).  The mean differences are listed
in Table~\ref{tab:alphadiffs}.  By this measure, the spectral
flattening between high-emissivity and deceleration regions is not
very significant ($2.5\sigma$) and that between coasting and
recollimation regions is consistent with zero. The remaining
differences show significant flattening ($>$4.5$\sigma$).  The most
significant difference is between the high-emissivity and
recollimation regions, for which $\langle \alpha_{\rm he}-\alpha_{\rm
  recoll}\rangle = 0.067 \pm 0.006$, consistent with a
constant spectral-index change for all of the 15 jets with
measurements at these locations (Fig.~\ref{fig:alpha_alpha}c).  The
particularly good agreement between the integrated spectral indices
measured for the outer two regions in the same jet, despite the
evidence for slight local variations in spectral index associated with
total-intensity enhancements (Section~\ref{trans}) suggests that the
internal errors cannot be much greater than we assume.

The internal consistency of the measurements is extremely good: all of
the jets for which we have accurate spectral indices show the same
trend.  There are a few discrepant points but these all represent
faint counter-jets for which the errors in our spectral index
estimates are large, so the discrepancies are not significant
(Fig.~\ref{fig:alpha_alpha}).  The rms spectral indices predicted by
the combination of random and calibration errors, $\sigma_{\rm pred}$,
are also close to the observed values (Table~\ref{tab:means}).  More
formally, we find that our measurements and error model are consistent
with all jets having the same spectral indices at the standard
locations: this hypothesis is excluded only at the 73\%, 40\%, 11\%
and 74\% confidence levels by a $\chi^2$ test for the high-emissivity,
deceleration, coasting and recollimation regions, respectively. The
validity of these confidence levels is, however, almost entirely
dependent on the accuracy of our estimate of calibration error: if we
have overestimated this, then some intrinsic dispersion in spectral
index would be required, albeit limited to $<$0.03 -- 0.05\,rms even
for perfect data.

Fig.~\ref{fig:alpha_alpha} also shows strong correlations between the
spectral indices measured in the same jet at different locations. The
significance levels of these correlations, derived from the Spearman
rank test, are given in Table~\ref{tab:alphadiffs}. Four of the
correlations are significant at $>$99.7\% confidence; the exceptions
are those between $\alpha_{\rm he}$ and $\alpha_{\rm coast}$ or
$\alpha_{\rm decel}$.  There are two possibilities: the correlations
result from residual calibration errors or the spectral-index evolution of
the jets follows a roughly similar trend, but with source-dependent
offsets. These cases are not mutually exclusive 
and are difficult to
distinguish, but we note that our assumed 3\% calibration error corresponds
to an rms spread in spectral index of $\approx$0.03 between 4.9 and
1.4\,GHz. This is very close to the rms spread along the line of
equality in any of the panels of Fig.~\ref{fig:alpha_alpha}.  If our
assumed calibration errors are accurate, then we cannot exclude the
possibility that all of the jets have identical spectral indices at
the fiducial locations.

\subsection{Spectral differences between main and counter-jets}

We find no statistically significant differences between the spectral
indices of the main and counter-jets at any of the four fiducial
locations.  The individual values are plotted in Fig.~\ref{fig:mj_cj}
and the mean differences and their errors are given in
Table~\ref{tab:mj_cj}.  As noted earlier, the only source which shows
a consistent offset between the two spectral indices is M\,84
(Fig.~\ref{fig:binnedprofiles}h), but this is almost certainly due to
imaging artefacts (Appendix~\ref{notes}).  The correlations between
spectral indices for the main and counter-jets at the fiducial
locations are not significant ($\leq$68\% using the Spearman rank
test; Table~\ref{tab:mj_cj}).

If the spectrum is not a pure power law and the jets are relativistic,
then systematic differences are expected between observed spectral
indices of approaching and receding jets.  We return to this point in
Section~\ref{constraints}.

\subsection{Transverse and local spectral-index variations}
\label{trans}

In detailed studies of 3C\,31, NGC\,315 and 3C\,296
\citep{LCBH06,LCCB06,3c31ls}, we drew attention to two additional
features of the spectral-index distributions.

Firstly, the spectrum of the main jet in NGC\,315 flattens away from
its axis \citep{LCCB06} as well as with distance from the AGN; the
same effect is visible in its counter-jet (Fig.~\ref{fig:images}b)
and, at a low level, on one side of each of the jets in 3C\,31
\citep[Fig.~\ref{fig:images}c of the present paper]{3c31ls}.  We have
found no similar systematic patterns in the rest of the sample.  It is
doubtful, however, that we could have reliably detected such a subtle
effect in sources with steeper-spectrum lobe emission surrounding the
jets given the simplicity of our linear subtraction algorithm
(Section~\ref{sub}).  Of the five sources where lobe emission is
undetectable over the region of interest, NGC\,315 and (at a much
lower level) 3C\,31 show the spectral flattening away from the axis
mentioned earlier, 0326+39 and 3C\,449 have no systematic gradients
and 1553+24 is not well enough resolved.  The pronounced transverse
spectral gradient in NGC\,315 therefore seems to be unusual, although
we cannot rule out similar effects at a lower level in other sources.
This gradient extends over many beam areas at high signal-to-noise
(Fig.~\ref{fig:images}b).  In contrast, the narrow rims of
steeper-spectrum emission seen close to the cores in NGC\,193, 3C\,31,
3C\,270 and 3C\,296 (Figs~\ref{fig:images}a, c, g, i) occur in regions
where the jets are narrow and faint, and are likely to result from
small calibration or deconvolution errors affecting the bright cores.
Narrow spectral-index features at the faint edges of jets at larger
distances from the nucleus are also suspect, particularly in cases
where significant lobe emission has been subtracted (e.g.\ the
counter-jet in 0206+35, Fig.~\ref{fig:images}d).

Secondly, we identified ``arcs'' -- narrow enhancements in total and
polarized intensity crossing the jets -- in 3C\,31 and 3C\,296. The
two brightest arcs in 3C\,31 have slightly ($\Delta\alpha \la 0.05$)
flatter spectra than the surrounding emission \citep{3c31ls}. In
3C\,296, the spectrum is flatter in both jets roughly where the more
distinct arcs are seen, but the arcs are not recognisable individually
on the spectral-index images \citep{LCBH06}.  The remaining
small-scale fluctuations visible in Fig.~\ref{fig:images} are
consistent with noise and calibration errors.

\section{Discussion}
\label{discuss}

\subsection{A common spectral-index profile?}
\label{alpha-profile}

\begin{figure}
\begin{center}
\includegraphics[width=8cm]{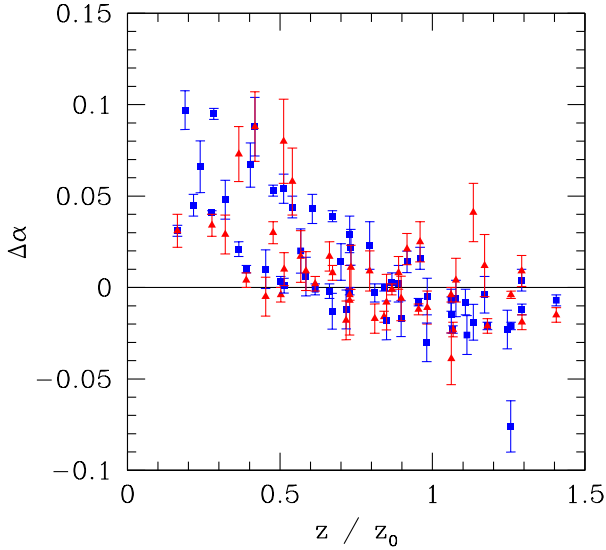}
\caption{Plot of $\Delta\alpha$ against normalized distance from the
  AGN.  $\Delta\alpha$ is the difference between the spectral
  index of a bin from Fig.~\ref{fig:binnedprofiles} and its mean value
  in the coasting region (or the recollimation region, if the
  former cannot be measured; see Table~\ref{tab:means}). Distance from
  the AGN, $z$, is normalized by the recollimation distance,
  $z_0$. Blue squares: main jets; red triangles: counter-jets. Data at $z/z_0 > 1.5$ are
  available only for a few jets and are not plotted. 0755+37
  (which has no measured reference spectral index) is also
  omitted, as are data from upstream of the flaring point in 3C\,270 and
  3C\,449.
\label{fig:plotallx}}
\end{center}
\end{figure}

Our jet models suggest that the flaring region is an approximately
homologous structure, in the sense that the width and fiducial
distances for emissivity and velocity all scale with the recollimation
distance despite the wide range in physical size (the recollimation
distance measured along the jet axis, $r_0$, ranges from 1.5\,kpc in
M\,84 to 35\,kpc in NGC\,315).  We will discuss these relations in detail
elsewhere (Laing \& Bridle, in preparation), but the scaling is
already evident from Table~\ref{tab:fiducials}.  The spectral indices
at our standard locations in the flow show very little dispersion between sources, so
they must also follow this scaling, as can be illustrated directly by
plotting the spectral index as a function of distance {\em normalized
  to the size of the flaring region}, $z/z_0$.  In order to remove the
confusing effects of scale and calibration errors, we evaluated the differences
$\Delta\alpha$ between the binned spectral indices in
Fig.~\ref{fig:binnedprofiles} and a reference value in the same jet.
We took the reference to be $\alpha_{\rm coast}$ ($\alpha_{\rm
  recoll}$ for 3C\,31 and 3C\,449, where the former is undefined),
forcing the mean profile for each jet to have $\Delta\alpha = 0$ at
the reference location.  The results are shown in
Fig.~\ref{fig:plotallx}.  There is a clear trend in the combined data:
on average, the spectral index falls from $\Delta\alpha \approx 0.06$
at $z/z_0 \approx 0.2$ (as close to the flaring point as we can
measure) to $\Delta\alpha \approx 0$ at $z/z_0 \approx 0.8$ (the
reference location).  This confirms our new result: {\em the
variations of spectral index appear to scale with the size of the
flaring region in the same way as variations of velocity and
emissivity.}

Jets evolve in cross-section and velocity as a consequence of mass
injection and propagation in external pressure and density gradients
(e.g.\ \citealt{LB02b}).  Their emissivities are in turn governed by
the balance of particle acceleration and energy-loss processes. The
most natural way to ensure a common scaling for these quantities is
for the acceleration mechanism to depend on flow speed, as we now
discuss.

\subsection{A relation between spectral index and flow speed?}
\label{alpha-vel}

\begin{figure}
\begin{center}
\includegraphics[width=8cm]{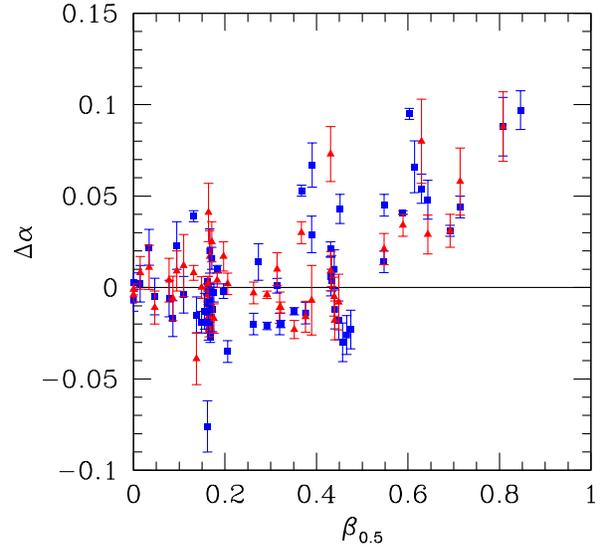}
\caption{Plot of $\Delta\alpha$ against jet flow
  velocity. $\Delta\alpha$ is the difference between the spectral
  index of a bin from Fig.~\ref{fig:binnedprofiles} and its mean value
  in the coasting region (or the recollimation region, if the former
  cannot be measured; see Table~\ref{tab:means}). The velocity
  $\beta_{0.5}$ is that for a streamline mid-way between the axis and
  the edge of the jet, from our model fits.  Blue squares: main jets;
  red triangles: counter-jets. We have not plotted the inner two bins
  for the main jet of 3C\,270, as these include significant emission
  from upstream of the flaring point.  0755+37 (no reference spectral
  index) and 3C\,449 (no velocity model) are also omitted.
\label{fig:plotallv}}
\end{center}
\end{figure}

Given that we model the jets as decelerating flows, and that the
spectral index decreases monotonically with distance from the AGN in
the regions studied here, there must be some relation between $\alpha$
and speed.  In order to quantify this relation, we have chosen a
representative speed $\beta_{0.5}$ for each source, determined from
our model for a streamline half way between the axis and boundary of
the jet. Spectral index difference (as in Section~\ref{alpha-profile})
is plotted against $\beta_{0.5}$ in Fig.~\ref{fig:plotallv}.  There is
a clear separation: all points with $\beta_{0.5} > 0.5$ have
$\Delta\alpha > 0$, whereas the mean $\Delta\alpha$ at lower velocity
is close to 0.

An upper bound 
to
the sound speed in the jet is that for an
ultrarelativistic plasma, $\beta_{\rm s} = 3^{-1/2} =
0.58$. $\beta_{\rm s}$ may be significantly lower than this limit,
particularly if the jet contains appreciable numbers of
non-relativistic protons (entrained or part of the initial
composition), and may change systematically with position in the jet
\citep{LB02b}. One possibility is then that the jet composition is
dominated by ultrarelativistic leptons and magnetic field with a minor
baryonic component and that the spectral index is significantly
higher than its asymptotic value of $\langle\alpha\rangle = 0.59$ 
wherever the average flow speed is even slightly supersonic. For the
10 sources we have modelled, the mean velocity at the flaring point is
$\beta = 0.81$ with an rms of 0.08. This in turn corresponds to a
generalized Mach number of ${\cal M} = (\Gamma\beta)/(\Gamma_{\rm
  s}\beta_{\rm s}) = 1.95$ for an ultrarelativistic plasma, where
$\Gamma_{\rm s} = (1-\beta_{\rm s}^2)^{-1/2}$.  For 3C\,31, we found
from a conservation-law analysis that the flow throughout the jet base
was transonic everywhere, with the Mach number decreasing from ${\cal
  M} \approx 2$ to ${\cal M} \approx 1$, despite the increasing mass
load \citep{LB02b}.  Any shocks in this flow regime must be weak.

An alternative possibility is that the 
sound speeds are much lower
than the ultrarelativistic limit, in which case larger Mach
numbers are possible for the flow.

\subsection{Constraints on particle acceleration}
\label{constraints}

We first estimate some fiducial numbers for magnetic fields, for the
radiating electron Lorentz factors and for energy loss timescales.
The on-axis magnetic-field strengths derived from our emissivity
models are $\approx 2 - 9$\,nT at the flaring point and $\approx 0.3 -
2$\,nT at recollimation\footnote{We assume that: the sum of energy
  densities of relativistic particles and magnetic field is a minimum
  (close to the equipartition condition); there are no relativistic
  protons; the filling factor is unity and the energy spectrum is a
  power law between electron Lorentz factors of 100 and $10^6$.}.  The
spectrum of synchrotron emission from a single electron is quite
broad, but a characteristic frequency is $\nu \approx \gamma^2
\nu_{\rm g}$, where $\gamma$ is the electron Lorentz factor and
$\nu_{\rm g} = eB/2\pi m_{\rm e}$ is the gyro frequency.  The corresponding 
Lorentz factor ranges for electrons radiating at 1.4 -- 8.5\,GHz are   
$2000 \la \gamma \la 10000$ at the flaring point and $5000 \la \gamma \la 30000$ at recollimation.
The synchrotron loss timescale is
\[
E/(-dE/dt) = \frac{3m_{\rm e}c\mu_0}{2\sigma_{\rm T}B^2\gamma}
\]
which ranges from $6 \times 10^5$ to $9 \times 10^7$\,yr for emission
at 8.5\,GHz.

A simple picture in which relativistic electrons are accelerated in
the high-emissivity region and then passively advected along the jet,
suffering radiative and adiabatic losses as they go, is not consistent
with the spectral flattening we observe: radiative losses cause the
spectrum to {\em steepen} at high energies.  Even if the radiating
particles are advected down the jet in a constant magnetic field, then
the observed frequency spectrum would appear to steepen with
distance. In fact, the magnetic field is likely to decrease away from
the AGN as a consequence of flux-freezing in a expanding flow. The
consequent rapid decrease may be partially compensated by
deceleration, shear or turbulent amplification, but (as mentioned
earlier) the equipartition field strength still falls by a factor of
$\approx$5 -- 10 between the flaring point and recollimation.  At a
fixed frequency, we therefore observe higher-energy electrons and the
steepening would be more pronounced.  In any case, the minimum loss
timescale for X-ray synchrotron emission is $\approx$100\,yr for our
assumed field strengths and a photon energy of 1\,keV. This
corresponds to a flow distance of $\approx 30/\beta$\,pc (much smaller
than the size of a typical jet base), so distributed reacceleration of
radiating electrons throughout the regions we observe in FR\,I jets is
unavoidable.  Independent evidence for ongoing acceleration comes from
modelling of radio brightness and polarization evolution along the
jets: additional electrons are also required to offset the effects of
adiabatic losses even at energies where the effects of synchrotron
cooling are small \citep{LB04}.  We must therefore attribute the
change in characteristic radio spectral index along the jets to the
underlying acceleration mechanism rather than to the combined effects
of synchrotron and adiabatic losses.  Although downward curvature at
high frequencies is seen in the broad-band spectra of jet bases
\citep{Evans05,H01,H05,Lanz,PW05,Perlman10,Worrall10,Harwood}, the
break frequencies are typically $\ga 10^{12}$\,Hz, well above the
range considered here.

We have measured the spatial variation of the slope of the radiation
spectrum over fixed ranges of observed frequency. In order to
characterize the particle-acceleration mechanism, we need to derive
the slope of the energy spectrum over a fixed (or, at least, known)
energy range.  Unless the spectrum is a perfect power law, this
presents several complications.
\begin{enumerate}
\item There is a known correction factor of $1+z$ for the source
  redshift. 
\item We infer relativistic flow speeds for the jets, so the ratio of
  observed and emitted frequencies ($\nu_{\rm obs}/\nu_{\rm em} =
  D$, where $D$ is the Doppler factor) may differ significantly from
  unity.  The emitted frequencies for the approaching and receding
  jets also differ systematically, but can be calculated from our jet
  models.
\item As mentioned earlier, the magnetic field strength is likely to decrease away
  from the AGN along the jet, so emission at a fixed frequency is
  generated by higher-energy electrons at larger distances.  The
  evolution of field strength with distance is subject to uncertain
  and mutually contradictory assumptions such as equipartition with
  relativistic particle energy density or flux freezing.
\end{enumerate}
The correction for redshift is small for these nearby sources 
(Table~\ref{tab:sources}) and it is
constant over both jets, so we can ignore it by comparison with the
other two effects, which we consider in turn.

The angles to the line of sight derived from our models range from
$\theta = 25^\circ$ to $\theta = 76^\circ$ and the maximum velocities
over the regions we measure spectral indices are in the range $0.75 <
\beta < 0.95$.  For 
constant-velocity, antiparallel flows, the
Doppler factors for the approaching and receding jets are:
\begin{eqnarray*}
D_{\rm j} & = & [\Gamma (1-\beta\cos\theta)]^{-1} \\
D_{\rm cj} & = & [\Gamma (1+\beta\cos\theta)]^{-1} \\
\end{eqnarray*}
If the energy spectrum deviates from a pure power law, but is the same
at a given distance from the AGN in both jets, then we would
expect a systematic dependence of spectral index difference
$\alpha_{\rm cj} - \alpha_{\rm j}$ on the ratio of Doppler factors
$D_{\rm j}/D_{\rm cj}$. The maximum effect for our data should occur
close to the flaring point, where the velocities are highest, and is
in principle very significant for our sample: the maximum predicted
ratio (for 1553+24) is $D_{\rm j}/D_{\rm cj} \approx 13$. Two problems
make the effect hard to observe: (a) the largest Doppler factor ratios
occur at small angles to the line of sight, so projection makes it
difficult to resolve the flaring point from the AGN and (b) if
$D_{\rm cj}$ is small, the counter-jet emission is faint and errors on
its spectral index are large. As a consequence, we have only been able to measure
spectral-index differences over the range $1 < D_{\rm j}/D_{\rm cj} < 
3.5$ for any of the fiducial regions, and those close to the AGN
are less well determined.  We see no systematic trend in plots of
$\alpha_{\rm cj} - \alpha_{\rm j}$ against $D_{\rm j}/D_{\rm cj}$ for
either the high-emissivity or coasting regions.

If all jets have the same electron energy spectrum which flattens with
increasing energy, then there is a natural explanation for the
observed spectral gradients, since the magnetic-field strength must
decrease away from the AGN.  It seems implausible, however, that
this (rather subtle) spectral flattening would occur in precisely the same 
{\em observed} frequency range for all of the jets in our sample, despite
the range of field strengths and Doppler factors. The flattening
cannot continue over a large energy range without violating
constraints from broad-band observations
(e.g.\ \citealt{H02,H05}). Finally, it is unclear what mechanism could
cause the flattening. This possibility should be tested using 
accurate measurements of jet spectra over a much larger
frequency range, but we consider it to be very unlikely.

The particle acceleration mechanism(s) at work in FR\,I jets must
therefore satisfy the following constraints:
\begin{enumerate}
\item The electron energy spectrum can be modelled as a pure power law
  with spatially-varying mean index $\langle p\rangle
  =2\langle\alpha\rangle+1$ restricted to the range $2.18 \leq \langle
  p\rangle \leq 2.32$ (Table~\ref{tab:means}). The characteristic
  electron Lorentz factor range probed by our observations is $2000 \la
  \gamma \la 30000$, given the assumptions made earlier.
\item The energy spectrum flattens with distance from the AGN,
  from $\langle p\rangle = 2.32$ in the high-emissivity region to
  $\langle p\rangle = 2.18$ in the coasting and
  recollimation regions.
\item By the argument of
  Section~\ref{alpha-vel} (Fig.~\ref{fig:plotallv}), $p$ may be
  determined by the characteristic bulk flow speed: a change in
  $\langle p\rangle$ from 2.32 to 2.18 being associated with a
  deceleration from $\beta \approx 0.8$ to $\beta \la 0.5$.
\item The mechanism responsible for the steeper energy spectrum in the
  high-emissivity regions must be capable of accelerating electrons to
  Lorentz factors of $\gamma \approx 10^8$, as demonstrated by the
  detection of X-ray emission in many cases
  (Table~\ref{tab:fiducials}, Fig.~\ref{fig:profiles}).  X-ray and/or
  optical emission is typically seen from the high-emissivity region
  of the brighter (approaching) jet; in some cases it extends slightly
  beyond the end of the high-emissivity region, but never beyond the
  end of deceleration. In the well-observed brighter jets
  of 3C\,31 and NGC\,315, the ratio of X-ray to radio emission
  decreases with distance from the AGN \citep{Girdwood}.  This
  decrease coincides with the spectral flattening we observe in the
  radio band and with deceleration.
\item The high-frequency spectrum at larger distances from the AGN
  (where we infer flatter energy spectra) is not known in most cases.
  The approaching jet of 3C\,31 is detected at 8\,$\mu$m wavelength out
  to a distance of $\approx$25\,arcsec \citep{Lanz}, implying that
  acceleration of electrons to $\gamma \approx 10^5 - 10^6$ (but not
  much higher) can occur after deceleration and recollimation.
\item Acceleration is distributed, rather than being associated with rare, 
  localized, prominent brightness enhancements.
\end{enumerate}

\subsection{Shock acceleration}
\label{shocks}

First-order Fermi acceleration is thought to be the dominant particle
acceleration mechanism at collisionless MHD shocks and there is an
extensive literature about this process (see \citealt{SB12} for a 
comprehensive list of references).  It is also the only process for which 
there are comprehensive predictions of the energy index under different physical
conditions, as we now discuss.

\subsubsection{Non-relativistic shocks}
\label{nonrel}

Early work (e.g.\ \citealt{Bell78,BO78}) established that a power-law
energy spectrum with index $p = 2$ ($\alpha = 0.5$) is produced by
strong, non-relativistic shocks in the test-particle approximation.
This value is inconsistent with our results.  Steeper spectra can in
principle be produced in weaker shocks: $p = (r+2)/(r-1)$, where $r$
is the velocity compression ratio, and $1 < r \leq 4$.  There is then
no obvious reason for $p$ to have a narrow range, however, as pointed out by 
\citet{Young}.

\subsubsection{The ultrarelativistic limit}
\label{ultra}

For ultrarelativistic parallel shocks (i.e.\ those in which the
magnetic field is parallel to the shock normal) in which particles
experience frequent small-angle scattering, a power-law energy
spectrum with $p = 2.23$ ($\alpha = 0.615$) is produced. This result
has been found both analytically \citep{Kirk00} and numerically
(e.g.\ \citealt{BO98,ED04,SB12}).  The coincidence between the
predicted spectral index and the value of 0.62 we found for three
sources in our earlier work \citep{Girdwood} led us to wonder whether
the ultrarelativistic limit might be relevant at the flaring point.
This was puzzling, as the asymptotic value is approached only for
shock Lorentz factors $\Gamma \ga 10$ and we infer $\Gamma \la 2$
downstream of the flaring point.  On the revised flux-density scale of
\citet{PB}, however, the inferred spectra are slightly but
significantly steeper, with $\langle\alpha\rangle = 0.66 \pm 0.01$ in
the high-emissivity region for the full sample.  Given this
discrepancy and the velocities $\beta \la 0.9$ that we infer
downstream of the flaring point in FR\,I jets, it now seems even less
likely that the ultrarelativistic limit is relevant to FR\,I jet
bases, although a spectral index of 0.62 is only marginally
inconsistent with our result if the rms error due to uncertainties in
the flux-density scale is as high as 0.02.  Closer to the nucleus, we
have no good estimates of the flow Lorentz factor, although
\citet{Perlman11} estimate $\Gamma \approx 4 - 5$ for the M\,87 jet at
knot HST-1.

\subsubsection{Mildly relativistic shocks}
\label{midrel}

The case of mildly relativistic shocks has been
addressed using Monte Carlo methods in the test-particle limit for
oblique shocks \citep{ED04,SB12} and in the non-linear case for
parallel shocks \citep{ED02}.  These studies show that mildly
relativistic shocks can generate a wide variety of power-law slopes
for the energy spectrum, depending on the obliquity of the field in
the shocks and the nature of the scattering as well as the shock
speed \citep{SB12}.  Our observations require a narrow range of energy
indices, with a slight dependence on mean flow (and therefore shock)
velocity.  The case which comes closest to meeting these requirements
is that of a mildly relativistic velocity upstream of the shock front
in the Bohm limit of frequent small-angle scatterings with a mean free
path $\lambda$ close to the electron gyro-radius $r_{\rm g}$
\citep[Figs~7 and 8]{SB12}.  The dependence of $p$ on field obliquity
is also weak under these circumstances.  The case of an upstream flow
velocity $\beta_{\rm u} = 0.71$, a velocity compression ratio of $r =
3.02$ and a sonic Mach number ${\cal M} = 2.6$ gives $2.35 \leq p \leq
2.5$, depending on obliquity \citep[Fig.~8]{SB12}; for $\beta_{\rm u}
= 0.1$ and $r = 4$, the range is $2 \leq p \leq 2.04$
\citep[Fig.~7]{SB12}, close to the limit of $p = 2$ for a
non-relativistic shock.  We suggest that these two cases might bracket
the range of physical conditions in FR\,I jet bases, the velocities in
the first case being slightly too fast for the flaring point, while those in
the second case are too slow for the flow after recollimation, which 
appears to remain significantly relativistic in most cases.  It is plausible that 
the upstream fluid velocities in the shock rest frame are in the appropriate 
range, for example if the shocks are stationary or moving slowly in the 
observed frame, and that there is a range of field obliquities.

\subsection{What are the particle acceleration mechanisms?}
\label{bestbuy}

The first-order Fermi process at mildly relativistic shocks in the
limit $\lambda \approx r_{\rm g}$ appears capable of generating energy
spectra with the correct indices and of accounting (at least
qualitatively) for their variations along the jets.  One possibility,
therefore, is that this is the sole acceleration mechanism.  A
potential difficulty, however, is the distributed nature of the
acceleration required to power the X-ray emission, which cannot be
restricted to a few localized shock sites: instead, a network of
shocks extending over a substantial distance along the jet appears to
be required.  Such shock networks could be identified with the
complex, non-axisymmetric brightness structure in the high-emissivity
region (e.g.\ Fig.~\ref{fig:sketch}d) or with the arcs observed
crossing the jets at larger distances in some sources
\citep{LCBH06,3c31ls}.  Efficient deceleration of the jet by
entrainment probably requires a transonic flow in which any shocks
would be very weak, however \citep{Bicknell94,LB02b}.

The most likely alternative to a pervasive shock network is 
that there are two distinct acceleration mechanisms operating in these
jets.
\begin{enumerate}
\item The first mechanism (which could again be diffusive shock
  acceleration) dominates wherever the flow velocity $\beta \ga 0.5$
  and in particular in the high-emissivity region.  It can accelerate
  electrons up to Lorentz factors $\gamma \approx 10^7 - 10^8$,
  enabling them to emit X-ray synchrotron radiation, and has a
  characteristic energy index $p = 2.32$ over the energy range we
  sample.
\item A second mechanism with a lower characteristic energy index $p
  = 2.18$ becomes important when the flow velocity falls below $\beta
  \approx 0.5$, and dominates after deceleration.  Gradual shear
  acceleration \citep{RD04,RD06} is a plausible candidate, as our jet
  models show that there are transverse velocity gradients in these
  regions. We originally favoured this idea for NGC\,315 on the
  grounds that the spectrum flattens away from the jet axis
  precisely where we infer velocity gradients \citep{LCCB06}, but this
  effect seems not to be general (Section~\ref{trans}).
  Second-order Fermi acceleration by plasma turbulence is a
  possible alternative mechanism. 
\end{enumerate}

The existence of two different acceleration mechanisms in FR\,I jet
bases has also been suggested based on evidence from high-resolution
observations of X-ray, optical and radio emission from the nearest
examples, Centaurus\,A \citep{H03,Goodger10} and M\,87
\citep{PW05,Perlman11}.  For Cen\,A, \citet{Goodger10} suggest that
stationary, X-ray-emitting knots are formed when lumps of dense
material (e.g.\ molecular clouds or stars with high mass-loss rates)
penetrate the jet, resulting in strong shocks with long-lived particle
acceleration up to high Lorentz factors.  Conversely, knots emitting
only in the radio band (which often have high speeds) would be formed
by weak shocks where no particles are accelerated to X-ray emitting
energies.  The proportion of radio emission from the two processes
changes systematically with distance from the AGN.  It is not known
whether there is a systematic difference in the radio spectra for the
two types of knot, as would be expected if the difference is related
to the spectral gradients we have discussed.

\section{Summary}
\label{summary}

The principal observational results from our imaging of jet base spectra 
for a sample of 11 FR\,I radio galaxies over the frequency range 
1.4 -- 8.5\,GHz
are as follows.
\begin{enumerate}
\item We find a narrow range of spectral indices, $\alpha$, with the
  mean values at our fiducial locations ranging from 0.66 to 0.59.
  The dispersion between sources is consistent with our error model.
\item The radio spectral indices decrease (i.e., the spectra flatten) with 
distance from the  nucleus on the scales we have studied. 
\item This spectral flattening occurs between the flaring point and
  the end of rapid deceleration
as determined from our modelling of the jets as decelerating
relativistic flows; thereafter the spectral index is
  constant until after the jets recollimate.
\item The spectral flattening is highly significant: the average decrease in
  spectral index between the high-emissivity region immediately
  downstream of the flaring point and just after the jet recollimates
  is $0.067 \pm 0.006$ and all jets for which we have data show
  spectral flattening between these two locations.
\item The mean spectral index for the high-emissivity region is
  $\langle\alpha\rangle = 0.662 \pm 0.011$; coasting and
  recollimation regions have $\langle\alpha\rangle = 0.592 \pm
  0.007$ and $0.594 \pm 0.008$, respectively.
\item There are no significant differences between the spectral
  indices of main and counter-jets at the same distance from the nucleus.
\item All of the jets show essentially the same variation of spectral
  index with distance from the nucleus when the distances are normalized 
  to the observed recollimation scale.
\item The steeper spectra close to the jet flaring points appear to be
  associated with bulk flow speeds $\beta \ga 0.5$.
\item The measured values of $\alpha$ correspond to indices of $p = 2.18 -
  2.32$ for the electron energy spectrum over a Lorentz factor range of
  2000 -- 30000, assuming equipartition magnetic fields.
\end{enumerate}
The range of observed spectral indices and the inferred dependence on
velocity could result from first-order Fermi acceleration by mildly
relativistic shocks, in the limit that the scattering mean free path
is close to the electron gyro-radius.  Such a model would require a
network of shocks in order to produce distributed particle
acceleration.

Alternatively, there may be two acceleration mechanisms. The first
can accelerate electrons to Lorentz factors of $10^7 - 10^8$, is 
dominant for bulk flow speeds $\beta \ga 0.5$ close to the flaring point and 
has p = 2.32 at low energies.  The
second dominates at slower flow speeds and thus at larger distances, 
has $p = 2.18$ and a lower maximum energy.  Shock acceleration is (again) a 
good candidate for the first mechanism

In order to distinguish between these (and other) alternatives,
accurate, well-sampled, spatially-resolved spectra over a very wide
frequency range (ideally from low-frequency radio to X-ray) are
required.  More detailed predictions of the energy spectra produced 
by various acceleration mechanisms, especially for Fermi acceleration by 
magnetized, mildly relativistic shocks in the Bohm limit, would also be 
valuable.

\section*{Acknowledgements}

We thank the referee, Eric Perlman, for a careful reading of the paper.

\appendix

\section{Archival VLA observations of 0326+39 and 1553+24}
\label{0326_1553}

The observations and data reduction for the images of 0326+39 and
1553+24 in the 4.5 -- 5.0 and 1.3 -- 1.7\,GHz bands have not been
documented elsewhere, so we summarize them here.  Data were retrieved
from the VLA archive, calibrated and imaged using the {\sc aips}
package, following standard procedures essentially as described by
\citet{LGBPB}.  3C\,48 or 3C\,286 were used as primary amplitude
calibrators.  All sources were self-calibrated and data from multiple
array configurations were combined. The final images were made using a
multi-scale clean algorithm \citep{MSC}.

For 1553+24, we chose not to make images at 1464.9\,MHz, as the
available A-configuration data at this frequency were badly
affected by interference.  We also analysed other archival 
datasets at 1665 and 1415\,MHz, but these were not observed in all
relevant configurations and did not help to improve the accuracy of
the final spectral-index image.
 
\begin{table}
\begin{minipage}{8.5cm}
\caption{Journal of VLA observations. (1) source name; 
  (2) date of observation; (3) VLA configuration (H denotes a
  hybrid configuration, as described by \citealt{Bridle91}); (4) and (5) 
  frequencies (MHz); (6) bandwidth (MHz); (7) duration (min); (8) 
  key to proposal code and reference.  \label{tab:obs0326+1553}}
\begin{tabular}{llllllrl}
\hline
&&&&&&&\\
0326+39 & 1980 04 15 & H & 4885.1 &        & 50 & 772 & 1\\
        & 1997 12 14 & D & 4885.1 & 4835.1 & 50 &  11 & 2\\
        & 1997 12 17 & D & 4885.1 & 4835.1 & 50 &  46 & 2\\
        & 1998 12 04 & C & 4885.1 & 4835.1 & 50 &  34 & 2\\
        & 1998 12 11 & C & 4885.1 & 4835.1 & 50 &  13 & 2\\
&&&&&&&\\
        & 1997 12 14 & D & 1385.1 & 1464.9 & 50 &  14 & 2\\
        & 1997 12 17 & D & 1385.1 & 1464.9 & 50 &  21 & 2\\
        & 1998 04 28 & A & 1385.1 & 1464.9 & 50 &  98 & 2\\
        & 1998 07 23 & B & 1385.1 & 1464.9 & 50 & 129 & 2\\
        & 1998 12 04 & C & 1414.9 & 1464.9 & 50 &  25 & 2\\
        & 1998 12 11 & C & 1385.1 & 1464.9 & 50 &   8 & 2\\
&&&&&&&\\
1553+24 & 1985 07 16 & C & 4885.1 & 4835.1 & 50 &  15 & 3\\
        & 1987 11 29 & B & 4885.1 & 4835.1 & 50 & 103 & 4\\
        & 1992 03 19 & C & 4885.1 & 4835.1 & 50 &  25 & 5\\
        & 1998 11 28 & C & 4885.1 & 4835.1 & 50 &  33 & 6\\
        & 1998 12 20 & C & 4885.1 & 4835.1 & 50 &  25 & 6\\
        & 1999 01 18 & C & 4885.1 & 4835.1 & 50 &  24 & 6\\
        & 1999 04 15 & D & 4885.1 & 4835.1 & 50 &  52 & 6\\
        & 2000 02 09 & B & 4885.1 & 4835.1 & 50 & 103 & 6\\
&&&&&&&\\
        & 1998 11 28 & C & 1385.1 &        & 50 &  15 & 6\\
        & 1998 12 20 & C & 1385.1 &        & 50 &  12 & 6\\
        & 1999 01 18 & C & 1385.1 &        & 50 &  12 & 6\\
        & 1999 09 20 & A & 1385.1 &        & 50 & 140 & 6\\
        & 2000 02 09 & B & 1385.1 &        & 50 &  64 & 6\\
&&&&&&&\\
\hline
\end{tabular}
Proposal codes and references:\\
1 BRID \citep{Bridle91}\\
2 AR386 (unpublished)\\
3 AM154 \citep{Morg87}\\
4 AM222 (unpublished)\\
5 AM364 \citep{Morg97}\\
6 AR402 \citep{Young}\\
\end{minipage}
\end{table}

\section{Notes on individual sources}
\label{notes}

\begin{description}
\item [{\bf NGC\,315}] A detailed study of the spectral-index distribution
  was given by \citet{LCCB06}.  The spectrum of the jets is
  unusually flat, although less extreme than in our earlier work after
  correction of the flux-density scale.
\item [{\bf 3C\,31}] A detailed study of the spectral index distribution was
  presented by \citet{3c31ls}.  The coasting region is not
  defined, as the deceleration region extends slightly beyond
  recollimation.
\item [{\bf 0206+35}] Spectral-index images of the whole source were
  shown by \citet{LGBPB} and for the jets alone (with a less stringent
  blanking criterion than used here) by \citet{LB12}. A residual
  calibration error caused a high-frequency ripple in spectral index
  transverse to the jet axis, visible in the brighter parts of the
  jets (Fig.~\ref{fig:profiles}d). The amplitude of this ripple is
  roughly $\Delta\alpha \approx 0.01$, and we do not believe that it
  has any systematic effect on our results.  The high-emissivity
  region is not well enough resolved from the core for us to determine
  accurate spectral indices.  We have included the part of the jet and
  counter-jet emission that we model as backflow \citep{LB12} in the
  binned spectral-index profiles and averages, since these are very
  hard to separate from the outflow component at 1.2-arcsec
  resolution.
\item [{\bf 0326+39}] \citet{Bridle91} show spectral-index profiles derived
  from lower-resolution images at 1.5 and 4.9\,GHz.
\item [{\bf 0755+37}] Spectral-index images of the whole source and
  for the jets have been presented elsewhere, as for 0206+35
  \citep{LGBPB,LB12}. Only the main jet has enough unblanked
  emission for us to derive a reliable spectral index. We considered
  using lower-resolution images to improve the signal-to-noise ration,
  but lobe subtraction was problematic. Although we model the jets as
  a mixture of outflow and backflow \citep{LB12}, the spectral indices
  in the present paper are dominated by the outflow component.
\item [{\bf 3C\,270}] A detailed discussion of the radio structure,
  spectral index and magnetic-field structure of 3C\,270 will be given
  by Laing, Guidetti \& Bridle (in preparation). The average spectral
  index of the jets of 3C\,270 is unusually high.  This is the only
  source where we can measure the spectral index accurately upstream
  of the flaring point. The abrupt steepening of the spectrum of the
  main jet from $\alpha \approx 0.55$ to $\alpha \approx 0.7$ between
  distances of $\approx$9 and $\approx$12\,arcsec from the nucleus is
  seen most clearly in the unbinned profile
  (Fig.~\ref{fig:profiles}g). Given that our model fits give an
  unusually large range for the flaring-point position (6.5 --
  11.5\,arcsec in projection), it is likely that the steepening occurs
  exactly at the flaring point. The location of the spectral
  steepening is blurred by the binning in
  Fig.~\ref{fig:binnedprofiles}(g).
\item [{\bf M\,84}] There are significant artefacts on the 1413-MHz image of
  M\,84, caused by a combination of poor spatial-frequency coverage,
  the proximity of the bright confusing source M\,87 and the
  displacement of the pointing centre in some of the observations
  \citep{LGBPB}. There are ripples of amplitude $\Delta\alpha \approx
  0.1$ and wavelength ranging from $\approx$3 to $\approx$15\,arcsec
  on the spectral-index image \citep[fig.~9c]{LGBPB}. These are
  aligned in the NS direction (roughly parallel to the jets).
\item [{\bf 3C\,296}] We have given a comprehensive discussion of the
  spectral-index distribution of 3C\,296 elsewhere
  \citep{LCBH06,LGBPB}.
\item [{\bf 1553+24}] \citet{Young} showed images of this source (we include
  some of their visibility data) but did not discuss its spectrum
  quantitatively. The high-emissivity region is too close to the core
  for accurate spectral-index measurements to be made; this is the one
  source where it does not overlap with the deceleration region.  The
  signal-to-noise ratio in the counter-jet is low.
\item [{\bf 3C\,449}] The spectra of the jets in this source were measured
  by \citet{Feretti99} and studied in detail by \citet{K-SR}. Our
  measured spectral indices are consistent with the mean value
  $\langle\alpha\rangle = 0.58 \pm 0.02$ between 4.985 and 8.4\,GHz
  with 2.5-arcsec resolution quoted by \citet{Feretti99} from the same
  visibility data once we account for recalibration. Note that the
  steep-spectrum emission surrounding the jets does not start until
  well outside our region of interest \citep{K-SR}. The inner jets are
  very symmetrical, indicating that they are almost precisely in the
  plane of the sky. For this reason, we cannot model the jet velocity
  field, and have no good estimate of the deceleration scale, although
  by analogy with other sources it is likely to be $\la$10\,arcsec.
\end{description}

\end{document}